\documentclass[sigconf, nonacm, pdfa]{acmart}

\newcommand\vldbdoi{10.14778/3681954.3682004}
\newcommand\vldbpages{3332 - 3345}
\newcommand\vldbvolume{17}
\newcommand\vldbissue{11}
\newcommand\vldbyear{2024}
\newcommand\vldbpagestyle{empty}
\newcommand\vldbauthors{\authors}
\newcommand\vldbtitle{\shorttitle} 
\newcommand\vldbavailabilityurl{https://github.com/michelleeesi/shapley_counterfactual}
\usepackage{color}
 \usepackage{setspace}
 \usepackage{balance}
 \usepackage{booktabs}
 \usepackage{amsmath}
 \usepackage{amsthm}
\usepackage{amsfonts}
\usepackage{graphics}
\usepackage{graphicx}
\usepackage{algorithmic}
\usepackage[a-2b]{pdfx}
[2018/12/22]
\usepackage[ruled,noend]{algorithm2e}
\usepackage{subcaption}
 \newtheorem{theorem}{Theorem}
 \newtheorem{property}{Property}
 
 \newtheorem{corollary}{Corollary}

\definecolor{mygreen}{rgb}{0,0.6,0}
\definecolor{mygray}{rgb}{0.5,0.5,0.5}
\definecolor{mymauve}{rgb}{0.58,0,0.82}
\usepackage{listings}
\usepackage{xcolor}
\usepackage{color, soul}
\newcommand{\todo}[1]{\textcolor{red}{\hl{[#1]}}}
\newcommand{\nop}[1]{}
\newcommand{\re}[1]{\textcolor{blue}{#1}}
\newcommand{\rem}[1]{\textcolor{purple}{#1}}

\SetCommentSty{mycommfont}

\SetKwInput{KwInput}{Input}                % Set the Input
\SetKwInput{KwOutput}{Output}              % set the Output
 
\definecolor{codegreen}{rgb}{0,0.6,0}
\definecolor{codegray}{rgb}{0.5,0.5,0.5}
\definecolor{codepurple}{rgb}{0.58,0,0.82}
\definecolor{backcolour}{rgb}{0.95,0.95,0.92}

\begin{document}
\begin{sloppy}
\title{Counterfactual Explanation of Shapley Value in Data Coalitions}

%%
%% The "author" command and its associated commands are used to define the authors and their affiliations.
\author{Michelle Si}
\affiliation{%
  \institution{Duke University}
  \streetaddress{P.O. Box 1212}
  \city{Durham, NC}
  \state{USA}
  \postcode{43017-6221}
}
\email{michelle.si@duke.edu}

\author{Jian Pei}
\orcid{0000-0002-1825-0097}
\affiliation{%
  \institution{Duke University}
  \streetaddress{1 Th{\o}rv{\"a}ld Circle}
  \city{Durham, NC}
  \country{USA}
}
\email{j.pei@duke.edu}

%%
%% The abstract is a short summary of the work to be presented in the
%% article.
\begin{abstract}
The Shapley value is widely used for data valuation in data markets. However, explaining the Shapley value of an owner in a data coalition is an unexplored and challenging task. To tackle this, we formulate the problem of finding the counterfactual explanation of Shapley value in data coalitions. Essentially, given two data owners $A$ and $B$ such that $A$ has a higher Shapley value than $B$, a counterfactual explanation is a smallest subset of data entries in $A$ such that transferring the subset from $A$ to $B$ makes the Shapley value of $A$ less than that of $B$. We show that counterfactual explanations always exist, but finding an exact counterfactual explanation is NP-hard. Using Monte Carlo estimation to approximate counterfactual explanations directly according to the definition is still very costly, since we have to estimate the Shapley values of owners $A$ and $B$ after each possible subset shift. We develop a series of heuristic techniques to speed up computation by estimating differential Shapley values, computing the power of singular data entries, and shifting subsets greedily, culminating in the SV-Exp algorithm. Our experimental results on real datasets clearly demonstrate the efficiency of our method and the effectiveness of counterfactuals in interpreting the Shapley value of an owner. 
\end{abstract}

\maketitle

%%% do not modify the following VLDB block %%
%%% VLDB block start %%%
\pagestyle{\vldbpagestyle}
\begingroup\small\noindent\raggedright\textbf{PVLDB Reference Format:}\\
\vldbauthors. \vldbtitle. PVLDB, \vldbvolume(\vldbissue): \vldbpages, \vldbyear.\\
\href{https://doi.org/\vldbdoi}{doi:\vldbdoi}
\endgroup
\begingroup
\renewcommand\thefootnote{}\footnote{\noindent
This work is licensed under the Creative Commons BY-NC-ND 4.0 International License. Visit \url{https://creativecommons.org/licenses/by-nc-nd/4.0/} to view a copy of this license. For any use beyond those covered by this license, obtain permission by emailing \href{mailto:info@vldb.org}{info@vldb.org}. Copyright is held by the owner/author(s). Publication rights licensed to the VLDB Endowment. \\
\raggedright Proceedings of the VLDB Endowment, Vol. \vldbvolume, No. \vldbissue\ %
ISSN 2150-8097. \\
\href{https://doi.org/\vldbdoi}{doi:\vldbdoi} \\
}\addtocounter{footnote}{-1}\endgroup
%%% VLDB block end %%%

%%% do not modify the following VLDB block %%
%%% VLDB block start %%%
\ifdefempty{\vldbavailabilityurl}{}{
\vspace{.3cm}
\begingroup\small\noindent\raggedright\textbf{PVLDB Artifact Availability:}\\
The source code, data, and/or other artifacts have been made available at \url{https://github.com/michelleeesi/shapley_counterfactual}.
\endgroup
}
%%% VLDB block end %%%

\section{Introduction}\label{sec:intro}

The power of big data largely comes from many secondary uses of data, such as enabling numerous machine learning and AI models~\cite{jordan2015machine, lecun2015deep, goodfellow2016deep}, recommender systems~\cite{resnick1997recommender, aggarwal2016recommender}, causal inference~\cite{hernan2010causal, pearl2010causal, pearl2009causal}, and data-driven decision-making applications~\cite{provost2013data}. However, incentivizing and facilitating data sharing and collaboration at scale remains a great challenge. Data markets~\cite{10.1145/2481528.2481532, 6691691, DBLP:conf/sigmod/ChenK019, 10.1145/3572751.3572757, 10.14778/3407790.3407800} are emerging as a promising way to enable and facilitate data sharing among many potential owners and consumers.  Essentially, a \textbf{data market} is an online platform where various parties with demands can acquire datasets or data services; at the same time, data owners can exchange data and data services for revenue or compensations in one way or another. There are already many active data markets, such as AWS Data Exchange, Windows Azure Marketplace, Dawex, Datarade, dmi.io, WorldQuant, Xignite, and BlueTalon~\cite{10.1145/3572751.3572757}.
\nop{
There are many different types of data markets due to many different needs in business models, vertical domains, government regulations, industry applications, and more.  %Designing, implementing, and deploying data markets in practice are highly interdisciplinary, involving techniques and expertises from cloud computing, secure storage and communication, databases, economics, law, business administration, marketing, and many other domains. 
In the realm of data market models and designs, the prospect of a one-size-fits-all solution appears improbable. Instead, principles and key modules ought to be crafted in a modular fashion, enabling their selection and assembly for the construction of distinct data markets tailored to various applications.
%Specific to the area of information integration and informatics, a focus is to develop the key techniques and components of data markets as information systems storing and integrating data.
}

At the core of every data market, there is a data valuation module.  In a data market, a group of data owners collaborate to produce a target dataset or complete a target task that a buyer would like to acquire, such as assembling a dataset for data analytics or machine learning.  We call collaboration among data owners a \textbf{data coalition}, where multiple datasets owned by different parties are used to produce target datasets. In a nutshell, \textbf{data valuation} assigns a score to a data owner to reflect the data owner's contribution towards a task achieved by a data coalition. 

Data valuation plays a central role in ensuring fairness, effectiveness, and efficiency of data markets.  There are various requirements on data valuation in different data markets~\cite{9300226}, such as truthfulness~\cite{10.1145/3328526.3329589}, revenue maximization, fairness~\cite{Shapley, 10.1145/3328526.3329589}, arbitrage-freeness~\cite{10.1145/2691190.2691191}, privacy-preservation~\cite{10.1257/jel.54.2.442, 10.1145/2229012.2229054, 10.1145/3219819.3220013, 10.5555/3241094.3241142, eurocrypt-2001-2009, 10.14778/3229863.3236266, DBLP:journals/corr/abs-1802-04780, 10.1145/2554797.2554835, 10.1145/2600057.2602902, 10.1145/1749603.1749605, Aggarwal2008, Bertino2008, 10.1007/978-3-540-79228-4_1, 10.1145/1540276.1540279, 7954609, Wu2010}, and computational efficiency~\cite{journals/pvldb/BalazinskaHS11, 10.1145/3328526.3329589, 10.5555/365411.365768}. The rich diversity in requirements poses many technical challenges for data valuation solutions.

The Shapley value~\cite{Shapley} is often used as a measure in data valuation~\cite{10.5555/3618408.3619157, pmlr-v151-kwon22a, wang2023threshold}, which is the expectation of marginal utility gain that a data owner can bring into coalitions. We will review the mathematical details in Section~\ref{sec:prob}. The Shapley value is the only valuation measure that provably satisfies efficiency, symmetry, zero element, and additivity.

While numerous studies have investigated the fast computation of Shapley values by either designing cost-saving sampling strategies~\cite{maleki2013bounding, pmlr-v89-jia19a} or tackling the Shapley computation in specific settings~\cite{10.1145/3514221.3517912}, one important question remains unexplored: \textbf{how can one understand and explain the Shapley value of a data owner in a data coalition?} 

Let us consider a concrete example. Suppose two data owners, Alice and Brittany, participate in a data coalition $\mathbb{O}$ and obtain their Shapley values $\psi_{\mathbb{O}}(Alice)$ and $\psi_{\mathbb{O}}(Brittany)$, respectively. Without loss of generality, let us assume $\psi_{\mathbb{O}}(Alice) > \psi_{\mathbb{O}}(Brittany)$. Then, one may ask how we can explain Alice's advantage over Brittany according to their Shapley values. To answer this question, one intuitive approach is to look for a counterfactual explanation, which is a minimal subset $S$ of data owned by Alice such that transferring $S$ from Alice to Brittany can flip the direction of the inequality, that is, making $\psi_{\mathbb{O}}(Alice)<\psi_{\mathbb{O}}(Brittany)$.

Using counterfactuals as an explanation tactic is a well established approach in philosophy and has enjoyed numerous applications in many domains~\cite{sep-counterfactuals}. A counterfactual explanation $S$ provides some interesting insights. For example, by checking the data entries in $S$, one may understand which data entries are the most crucial for Alice's advantage—what really makes Alice be able to contribute more and thus be more valuable than Brittany in the coalition?  
%The size of $S$ may indicate how easily Brittany can overcome Alice's advantage—if $|S|$ is small, then Brittany can take away Alice's advantage by obtaining a small number of data entries, but if $|S|$ is large, then Brittany might need to obtain a significantly large amount of data to outperform Alice. 
The counterfactual explanation allows us to detail the root of differences in Shapley values between data owners in a way that the Shapley value itself does not illuminate. For example, if Alice owns much more data than Alice but the size of the counterfactual explanation is small, it means that Alice's advantage is driven by a few powerful rows of data--the elements of the counterfactual. To use another analogy, using a counterfactual explanation would allow us to see if, given two sports teams, if the success of one over the other on average is driven by one star player (a small counterfactual) or an overall better playing team (a large counterfactual). In Section~\ref{sec:exp}, we report two interesting case studies. The first one demonstrates that counterfactual explanations can help select features from one subset to enhance the features in another subset. The second case illustrates that counterfactual explanations can disclose the data distribution differences among different data owners.

Motivated by this insight, we tackle the problem of finding the counterfactual explanation for the Shapley value. To the best of our knowledge, we are the first to model this problem.  We show that the counterfactual explanation always exists, but finding the counterfactual explanation of the Shapley value is NP-hard.

Even if we use Monte Carlo estimation to approximate counterfactual explanations according to the definition, the algorithm is still very costly--we have to estimate the Shapley values of data owners after every change. To address the computational challenges, we develop a series of heuristic techniques to improve computation. Firstly, we provide techniques to estimate the differential Shapley (the difference between the Shapley values of two data owners) directly. The differential Shapley not only avoids the complication of estimating the Shapley values of two data owners and then calculating the difference, but also improves the estimation quality when using Monte Carlo sampling approaches.  Secondly, estimating differential Shapley values is still costly when there are many data entries. For this, we develop an iterative greedy search approach. In each iteration, we find a data entry such that moving the data entry from one data owner to the other causes an estimated maximal change in differential Shapley value, which is measured using the notion of the \emph{power} of a data entry. The iteration continues until an approximation of the counterfactual explanation is obtained.

We also conduct experiments and highlight case studies on real datasets to examine the efficiency of our method and the effectiveness of using counterfactuals to explain the Shapley value.  %\rem{We focus on using the differential Shapley value for feature selection purposes to show the advantages of explanability in understanding feature importance}.

The rest of the paper is organized as follows.  We formulate the problem and present an exact algorithm in Section~\ref{sec:prob}.  In Section~\ref{sec:methods} we develop the heuristic approximation methods. We report the experimental results in Section~\ref{sec:exp}, discuss related work in Section~\ref{sec:related}, and address the limitations and possible extensions of our method in Section~\ref{sec:ext}. Section~\ref{sec:con} concludes the paper.

\section{Problem Formulation and an Exact Algorithm}\label{sec:prob}

Assume a set of data entries $\mathbb{D}=\{x_1, \ldots, x_m\}$ of interest.  Let $\mathbb{O}$ be a set of data owners who achieve a task in a coalition.  For each data owner $O \in \mathbb{O}$, we overload symbol $O$ to also denote the dataset that $O$ owns, that is, $O \subseteq \mathbb{D}$ is the subset of data entries that $O$ owns. A (data) \textbf{coalition} $\mathcal{S} \subseteq \mathbb{O}$ is a subset of data owners and their datasets.  Correspondingly, the union of datasets owned by a set of data owners $\mathcal{S}$, that is, $\cup_{O \in \mathcal{S}}O \subseteq \mathbb{D}$, is called a \textbf{composed dataset}.  $\mathbb{O}$ itself as a coalition is called the \textbf{grant coalition}.

Given a set of data owners $\mathbb{O}$ and a data owner $O \in \mathbb{O}$, denote by $\psi_\mathbb{O}(O)$ the Shapley value~\cite{Shapley} of $O$, that is,
\begin{equation}\label{eq:shapley}
  \begin{split}
  \psi_\mathbb{O}(O) & =\frac{1}{|\mathbb{O}|}\sum_{\mathcal{S} \subseteq \mathbb{O} \setminus \{O\}}\frac{U(\mathcal{S}\cup\{O\})-U(\mathcal{S})}{{{|\mathbb{O}|-1} \choose {|\mathcal{S}|}}}\\
  & =\frac{1}{|\mathbb{O}|!}\sum_{\pi \in \Pi(\mathbb{O})}(U(P^\pi_O \cup \{O\})-U(P^\pi_O))
  \end{split}
\end{equation}
where $U: 2^\mathbb{O} \rightarrow \mathbb{R}$ is a \textbf{utility function} that returns the utility of a coalition by a set of data owners, $\mathbb{R}$ is the set of real numbers, $U(\emptyset)=0$, $\Pi(\mathbb{O})$ is the set of all possible permutations of data owners, and $P^\pi_O$ is the set of data owners preceding $O$ in permutation $\pi$.

In many applications, the more data, the better the utility. In other words, a utility function is often monotonic.  Even with a non-monotonic utility function $U(\cdot)$, since a user often has the incentive to try every possible way to extract the best value from a set of data, the attempts lead to a utility function $U^*(D)=\max_{D' \subseteq D}\{ U(D')\}$, which is monotonic. Based on this rationale, we from now on assume that the utility function is \textbf{monotonic}.  That is, for any two subsets of data entries $D_1, D_2 \subseteq \mathbb{D}$, if $D_1 \subseteq D_2$, then $U(D_1) \leq U(D_2)$.

\nop{Once a user obtains a dataset, the user can use all ways to extract the best value from the dataset.  Therefore, it is reasonable to assume that the utility function is \textbf{monotonic}.  That is, for any two subsets of data entries $D_1, D_2 \subseteq \mathbb{D}$, if $D_1 \subseteq D_2$, then $U(D_1) \leq U(D_2)$. \todo{handwave monotonicity assumption - may not be true always but assume true for this problem}
}

Consider two data owners $A, B \in \mathbb{O}$ such that $\psi_\mathbb{O}(A)>\psi_\mathbb{O}(B)$. We ask the following question: \emph{which data entries in $A$ can explain the higher Shapley value of $A$ compared to that of $B$?}  Particularly, we are interested in the \textbf{counterfactual explanation}, that is, a subset $\Delta A \subseteq A$ of the minimum size such that, if $\Delta A$ is transferred from $A$ to $B$, the Shapley value of $B$ will be larger than that of $A$.  

Formally, let $\mathbb{O}[A\xrightarrow{\Delta A}B]=\mathbb{O}\setminus \{A, B\} \cup \{A \setminus \Delta A, B \cup \Delta A\}$. We want to solve the following \textbf{counterfactual explanation problem of the Shapley value} as an optimization problem.
\begin{equation}\label{eq:problem}
\begin{split}
    & \min\{|\Delta A|\} \\
    \text{s.t. } & \Delta A \subseteq A \\
    & \psi_{\mathbb{O}[A\xrightarrow{\Delta A}B]}(A \setminus \Delta A) < \psi_{\mathbb{O}[A\xrightarrow{\Delta A}B]}(B \cup \Delta A)
%    & \text{where } \mathbb{O'}= \mathbb{O}[A\xrightarrow{\Delta A}B]
  \end{split}
\end{equation}

We can show that the problem of counterfactual explanation is NP-hard. 

\begin{theorem}[Complexity]
The problem of counterfactual explanation is NP-hard.
\proof[Proof sketch]
We prove by constructing a reduction from the set cover problem, whose search version is known to be NP-hard~\cite{Karp1972}. Given a set of elements $\mathbb{D}=\{x_1, \ldots, x_n\}$ (called the universe) and a collection $\mathbb{S}=\{S_1, \ldots, S_m\}$ of $m$ subsets whose union equals the universe, that is, $S_i \subseteq \mathbb{D}$ and $\cup_{i=1}^m S_i = \mathbb{D}$, the set cover problem is to find the smallest sub-collection of $\mathbb{S}$ whose union equals the universe $\mathbb{D}$.

For each sub-collection $\mathcal{S}=\{S_{i_1}, \ldots, S_{i_k}\} \subseteq \mathbb{S}$, we define an encoding function $f(\mathcal{S})=\frac{\sum_{j=1}^k 2^{i_j}}{2^{m+1}}$. Clearly, $0 < f(\mathcal{S}) < 1$ as long as $\mathcal{S} \neq \emptyset$.

We construct a game as follows. We treat each subset $S_i$ as a data entry. There are only two data owners, $O_0=\emptyset$ and $O_1=\mathbb{S}$, that is, $O_0$ does not have any data and $O_1$ has all the subsets. We define a utility function $U:2^\mathbb{S} \rightarrow [0, m+1)$ such that for each sub-collection $\mathcal{S}\subseteq \mathbb{S}$, $U(\mathcal{S})=0$ if $\cup_{S_i \in \mathcal{S}} S_i \neq \mathbb{D}$; and otherwise $U(\mathcal{S})=m-|\mathcal{S}|+f(\mathcal{S})$.

%$U(\mathcal{S})=1$ if $\cup_{S_i \in \mathcal{S}} O_i = \mathbb{D}$; otherwise $U(\mathcal{S})=0$.

Clearly, $\psi(O_1)>0$ and $\psi(O_0)=0$, and thus $\psi(O_1) > \psi(O_0)$. Let $\Delta O \subseteq O_1$ be a counterfactual explanation of the Shapley value. Then, one of the following three cases may happen. Firstly, if $\mathbb{S}$ has only one cover $\mathcal{S}\subseteq \mathbb{S}$, then $\Delta O=\mathcal{S}$. Secondly, if $\mathbb{S}$ has two or more covers and the minimal cover $\mathcal{S}$ satisfies $|\mathcal{S}| < |\mathbb{S} \setminus \mathcal{S}|$, then $\Delta O=\mathcal{S}$ (note that if the size of a cover $S$ is strictly greater than $|\mathbb{S} \setminus \mathcal{S}|$, then $\mathcal{S}$ cannot be minimal). Otherwise, we have the third case, where there are two covers and the minimal cover $\mathcal{S}$ satisfies $|\mathcal{S}| = |\mathbb{S} \setminus \mathcal{S}|$. Thus, $\mathbb{S}$ has two disjoint covers $\mathcal{S}_1$ and $\mathcal{S}_2$ such that $\mathbb{S}=\mathcal{S}_1 \cup \mathcal{S}_2$ and $\mathcal{S}_1 \cap \mathcal{S}_2=\emptyset$, and then we have that $\Delta O=\arg\max_{\mathcal{S}\in\{\mathcal{S}_1, \mathcal{S}_2\}}U(\mathcal{S})$.
Therefore, the subsets in $\Delta O$ is a set cover in $\mathbb{D}$. 
% \re{Please note that, if the size of a cover $S$ is strictly greater than $|\mathbb{S} \setminus \mathcal{S}|$, then $S$ cannot be minimal.}
\qed
\end{theorem}

It is important to note the feasibility of this problem.

\begin{proposition}\label{cor:feasibility}
The counterfactual explanation problem of Shapley value always has a feasible solution. 

\proof[Proof sketch]
Since $\psi_\mathbb{O}(A) > \psi_\mathbb{O}(B)$ and the utility function is monotonic, the Shapley value is non-negative, that is, $\psi_\mathbb{O}(A) >0$.  Therefore, $A \neq \emptyset$.  Let $\mathbb{O'}=\mathbb{O}[A\xrightarrow{\Delta A}B]$. In the trivial scenario where $\Delta A = A$, due to the monotonicity of the utility function, we have $\psi_\mathbb{O'}(A \setminus \Delta A) =\psi_\mathbb{O'}(\emptyset)=0$. Moreover, $B \cup \Delta A \supseteq \Delta A =A \neq \emptyset$.  Thus, $\psi_\mathbb{O'}(B \cup \Delta A) >0$.  Then, $\psi_\mathbb{O'}(A \setminus \Delta A)< \psi_\mathbb{O'}(B \cup \Delta A)$.  The feasibility of the problem follows immediately.
\qed
\end{proposition}

We now introduce the notion of the \textbf{differential Shapley value}. For two data owners $A, B \in \mathbb{O}$, we define the differential Shapley value between $A$ and $B$ as $\Psi_\mathbb{O}(A, B)=\psi_\mathbb{O}(A)-\psi_\mathbb{O}(B)$.  We have the following useful result, a variation of Lemma 1 by~\citet{pmlr-v89-jia19a} on efficient data valuation.

\begin{theorem}[Differential Shapley Value]\label{thm:differential} For two data owners $A, B \in \mathbb{O}$,
\[ 
\Psi_\mathbb{O}(A, B)= \sum_{\mathcal{S} \subseteq \mathbb{O} \setminus \{A \cup B\}} \frac{1}{(|\mathcal{S}|+1) {{|\mathbb{O}|-1} \choose {|\mathcal{S}|+1}}} (U(\mathcal{S}\cup \{A\})-U(\mathcal{S}\cup \{B\}))
\]
\proof According to the definition of the Shapley value (Equation~\ref{eq:shapley}), we have
%\scriptsize
\begin{align*}
\psi_\mathbb{O}(A)=&\frac{1}{|\mathbb{O}|} \sum_{\mathcal{S} \subseteq \mathbb{O} \setminus \{A\}}\frac{U(\mathcal{S}\cup\{A\})-U(\mathcal{S})}{{{|\mathbb{O}|-1} \choose {|\mathcal{S}|}}}\\
=& \frac{1}{|\mathbb{O}|}\sum_{\mathcal{S} \subseteq \mathbb{O} \setminus \{A \cup B\}}\frac{U(\mathcal{S}\cup\{A\})-U(\mathcal{S})}{{{|\mathbb{O}|-1} \choose {|\mathcal{S}|}}} \\
&+\frac{1}{|\mathbb{O}|}\sum_{\mathcal{S} \subseteq \mathbb{O} \setminus \{A \cup B\}}\frac{U((\mathcal{S}\cup \{B\})\cup\{A\})-U(\mathcal{S}\cup \{B\})}{{{|\mathbb{O}|-1} \choose {|\mathcal{S}|+1}}}
\end{align*}
%\normalsize

Similarly,
%\scriptsize
\begin{align*}
\psi_\mathbb{O}(B)=&\frac{1}{|\mathbb{O}|}\sum_{\mathcal{S} \subseteq \mathbb{O} \setminus \{A \cup B\}}\frac{U(\mathcal{S}\cup\{B\})-U(\mathcal{S})}{{{|\mathbb{O}|-1} \choose {|\mathcal{S}|}}}\\
&+\frac{1}{|\mathbb{O}|}\sum_{\mathcal{S} \subseteq \mathbb{O} \setminus \{A \cup B\}}\frac{U((\mathcal{S}\cup \{A\})\cup\{B\})-U(\mathcal{S}\cup \{A\})}{{{|\mathbb{O}|-1} \choose {|\mathcal{S}|+1}}}
\end{align*}
%\normalsize
Thus,
\begin{align*}
& \Psi_\mathbb{O}(A, B)=\psi_\mathbb{O}(A)-\psi_\mathbb{O}(B)\\
=&
\frac{1}{|\mathbb{O}|}\sum_{\mathcal{S} \subseteq \mathbb{O} \setminus \{A \cup B\}}\frac{U(\mathcal{S}\cup\{A\})}{{{|\mathbb{O}|-1} \choose {|\mathcal{S}|}}}-\frac{1}{|\mathbb{O}|}\sum_{\mathcal{S} \subseteq \mathbb{O} \setminus \{A \cup B\}}\frac{U(\mathcal{S}\cup \{B\})}{{{|\mathbb{O}|-1} \choose {|\mathcal{S}|+1}}}\\
&-\frac{1}{|\mathbb{O}|}\sum_{\mathcal{S} \subseteq \mathbb{O} \setminus \{A \cup B\}}\frac{U(\mathcal{S}\cup\{B\})}{{{|\mathbb{O}|-1} \choose {|\mathcal{S}|}}}+\frac{1}{|\mathbb{O}|}\sum_{\mathcal{S} \subseteq \mathbb{O} \setminus \{A \cup B\}}\frac{U(\mathcal{S}\cup \{A\})}{{{|\mathbb{O}|-1} \choose {|\mathcal{S}|+1}}}\\
% =& \frac{1}{|\mathbb{O}|}\left(\sum_{\mathcal{S} \subseteq \mathbb{O} \setminus \{A \cup B\}}\frac{U(\mathcal{S}\cup\{A\})}{{{|\mathbb{O}|-1} \choose {|\mathcal{S}|}}} + \sum_{\mathcal{S} \subseteq \mathbb{O} \setminus \{A \cup B\}}\frac{U(\mathcal{S}\cup \{A\})}{{{|\mathbb{O}|-1} \choose {|\mathcal{S}|+1}}} \right)\\
% & - \frac{1}{|\mathbb{O}|}\left(\sum_{\mathcal{S} \subseteq \mathbb{O} \setminus \{A \cup B\}}\frac{U(\mathcal{S}\cup\{B\})}{{{|\mathbb{O}|-1} \choose {|\mathcal{S}|}}} + \sum_{\mathcal{S} \subseteq \mathbb{O} \setminus \{A \cup B\}}\frac{U(\mathcal{S}\cup \{B\})}{{{|\mathbb{O}|-1} \choose {|\mathcal{S}|+1}}} \right)\\
=& \frac{1}{|\mathbb{O}|}\sum_{\mathcal{S} \subseteq \mathbb{O} \setminus \{A \cup B\}} \frac{\left({{|\mathbb{O}|-1} \choose {|\mathcal{S}|+1}} + {{|\mathbb{O}|-1} \choose {|\mathcal{S}|}} \right)}{{{|\mathbb{O}|-1} \choose {|\mathcal{S}|+1}} {{|\mathbb{O}|-1} \choose {|\mathcal{S}|}}} (U(\mathcal{S}\cup \{A\})-U(\mathcal{S}\cup \{B\}))\\
% =& \frac{1}{|\mathbb{O}|}\sum_{\mathcal{S} \subseteq \mathbb{O} \setminus \{A \cup B\}} \frac{|\mathbb{O}|}{(|\mathcal{S}|+1) {{|\mathbb{O}|-1} \choose {|\mathcal{S}|+1}}} (U(\mathcal{S}\cup \{A\})-U(\mathcal{S}\cup \{B\}))\\
=& \sum_{\mathcal{S} \subseteq \mathbb{O} \setminus \{A \cup B\}} \frac{1}{(|\mathcal{S}|+1) {{|\mathbb{O}|-1} \choose {|\mathcal{S}|+1}}} (U(\mathcal{S}\cup \{A\})-U(\mathcal{S}\cup \{B\})) \qed
\end{align*}
\end{theorem}

Since computing the exact Shapley value is \#P-hard~\cite{doi:10.1287/moor.19.2.257}, Theorem~\ref{thm:differential} allows us to work on the difference between the Shapley values of two data owners directly without estimating the individual values. 

%This result is important because it allows us to use a weaker definition of confidence in our approximation algorithms, which results in faster computation time. Approximating the actual Shapleys of two different data owners up to a certain $\delta$-confidence interval is an ambiguous task, as the question of when to stop sampling is not entirely clear. The differential Shapley, however, is perfectly suited to answer the question of when two data owners' Shapley's have flipped. We are only interested in knowing if the differential Shapley is positive ($A$ has the higher Shapley) or negative ($B$ has the higher Shapley), provide a clear null hypothesis ($\psi_\mathbb{O}(A) - \psi_\mathbb{O}(B)\geq0$) against which to test. We can now define $\delta$-confidence as the confidence that our differential Shapley is not greater than or equal to zero. 

To obtain the exact counterfactual explanation of Shapley value, a straightforward approach uses Theorem~\ref{thm:differential}, enumerates all possible non-empty subsets $\Delta A$ of $A$, and checks whether the resulting datasets satisfy Equation~\ref{eq:problem}. Among all those $\Delta A$ satisfying Equation~\ref{eq:problem}, we pick the one with the smallest size. Since we are searching for the smallest subset, we can enumerate the subsets of $A$ in the size-ascending order. The algorithm stops the first time Equation~\ref{eq:problem} is satisfied, guaranteeing that $\Delta A$ is minimum in size.  The pseudocode is given in Algorithm~\ref{algo:bruteforce}. Clearly, a straightforward implementation of Algorithm~\ref{algo:bruteforce} generates the powerset of $A$ in the worst case, and thus its time complexity is exponential.

%Additionally, the Shapley value calculation involves a large number of summations of the utility function outputs. The utility function could be very complicated, and thus this naive greedy method would be extremely inefficient.\\%

\begin{algorithm}[t]
\DontPrintSemicolon
%\setstretch{1.3}
\SetAlgoLined
\KwInput{A set of data owners $\mathbb O$ and two data owners $A, B \in \mathbb O$ such that $\psi_{\mathbb O}(A) > \psi_{\mathbb O}(B)$
%, a utility matrix $[W]_{|{\mathbb O}| \times |{\mathbb O}|}$
  }
\KwOutput{Solution to Equation~\ref{eq:problem}}
\For{$i=1$ \textbf{\em to} $|A|-1$}{
    \For {$\Delta A \subset A$ s.t. $|\Delta A|=i$}{
    Let $\mathbb{O'}= \mathbb{O}\setminus \{A, B\} \cup \{A \setminus \Delta A, B \cup \Delta A\}$;\\
    \If{$\psi_\mathbb{O'}(A \setminus \Delta A) < \psi_\mathbb{O'}(B \cup \Delta A)$}
    {
       \Return{$\Delta A$
    }
  }
}
}
\caption{The Brute-force Exact Algorithm \label{algo:bruteforce}}
\end{algorithm}

\section{Heuristic Approximation Methods}\label{sec:methods}

Computing the exact answer to the counterfactual explanation problem is costly and does not scale up for the scenarios where there are many data owners.  Thus, in this section we explore heuristic methods in two ways.

Firstly, we approximate the difference between the Shapley values of two data owners through Monte Carlo sampling. Secondly, we explore how to estimate Shapley value changes when we move some data entries from one data owner to the other and then greedily search for a counterfactual explanation.  

\subsection{Estimating Differential Shapley Value}\label{sec:mc-baseline}

A straightforward approach to estimate the difference between the Shapley values of two data owners is to first estimate the individual Shapley values and then infer the difference. This approach has two drawbacks. Firstly, we need to draw samples to estimate the two individual Shapley values, and each has independent estimation error. Secondly, the estimation of the difference using the estimated individual Shapley values introduces a new estimation error. Can we reduce the estimation error and improve efficiency by estimating the difference directly?

According to Theorem~\ref{thm:differential}, the differential Shapley value between two data owners depends only on the difference between the marginal utility $U(\mathcal{S} \cup \{A\})$ and $U(\mathcal{S} \cup \{B\})$ on all coalitions $\mathcal{S}$ in which neither $A$ nor $B$ participates. This insight can be used in estimating the differential Shapley value directly using Monte Carlo approximation.

\begin{corollary}[Differential Shapley Value by Permutation]\label{cor:mc-baseline}
For two data owners $A, B \in \mathbb{O}$,
\[
\begin{split}
%& \psi_\mathbb{O}(A) - \psi_\mathbb{O}(B)\\
\Psi_\mathbb{O}(A, B) =& \frac{1}{2(|\mathbb{O}|-1)!}\sum_{\pi \in \Pi(\mathbb{O})} \frac{1}{|\mathbb{O}|-|P^\pi_{\{A, B\}}|-1}(U(P^\pi_{\{A, B\}} \cup \{A\})\\
&-U(P^\pi_{\{A, B\}} \cup \{B\}))
\end{split}
\]
where 
%$\Pi(\mathbb{O})$ is the set of all possible permutations of data owners except for $A$ and $B$, and 
$P^\pi_{\{A, B\}}$ is the set of data owners preceding $A$ and $B$ in permutation $\pi$.
\proof
For each subset of data owners $\mathcal{S} \subseteq \mathbb{O} \setminus \{A, B\}$, there are 
$|\mathcal{S}|! \times 2 \times (|\mathbb{O} \setminus \mathcal{S}|-1)!=2|\mathcal{S}|!(|\mathbb{O}|-|\mathcal{S}|-1)!
$
permutations $\pi$ in $\Pi(\mathbb{O})$ such that $P^\pi_{\{A, B\}}=\mathcal{S}$.  Following Theorem~\ref{thm:differential}, we have  
\[
\begin{split}
%& \psi_\mathbb{O}(A) - \psi_\mathbb{O}(B)\\
\Psi_\mathbb{O}(A, B)=& \sum_{\mathcal{S} \subseteq \mathbb{O} \setminus \{A \cup B\}} \frac{1}{(|\mathcal{S}|+1) {{|\mathbb{O}|-1} \choose {|\mathcal{S}|+1}}} (U(\mathcal{S}\cup \{A\})-U(\mathcal{S}\cup \{B\}))\\
=& \sum_{\pi \in \Pi(\mathbb{O})} \frac{1}{(|P^\pi_{\{A, B\}}|+1) {{|\mathbb{O}|-1} \choose {|P^\pi_{\{A, B\}}|+1}}}\\
& \times \frac{(U(P^\pi_{\{A, B\}} \cup \{A\})-U(P^\pi_{\{A, B\}} \cup \{B\}))}{2|P^\pi_{\{A, B\}}|!(|\mathbb{O}|-|P^\pi_{\{A, B\}}|-1)!}\\
=& \frac{1}{2(|\mathbb{O}|-1)!}\sum_{\pi \in \Pi(\mathbb{O})} \frac{1}{|\mathbb{O}|-|P^\pi_{\{A, B\}}|-1}(U(P^\pi_{\{A, B\}} \cup \{A\})\\
&-U(P^\pi_{\{A, B\}} \cup \{B\})) \qed
\end{split}
\]
\end{corollary}

The corollary immediately leads to a Monte Carlo estimator, which takes a uniform sample of permutations $X \subseteq \Pi(\mathbb{O})$ to estimate the differential Shapley value $\Psi_{\mathbb{O}}(A,B)$. We can show that this estimation is unbiased.

\begin{corollary}[Unbiased Estimate]\label{cor:diff-est}
For a uniform sample of permutations $X \subseteq \Pi(\mathbb{O})$,
\[
\begin{split}
%& \widehat{\psi_\mathbb{O}(A) - \psi_\mathbb{O}(B)}\\
& \Delta \psi_{\mathbb{O}}^{A-B}\\
=& \frac{|\mathbb{O}|}{2|X|}\sum_{\pi \in X} \frac{1}{|\mathbb{O}|-|P^\pi_{\{A, B\}}|-1}(U(P^\pi_{\{A, B\}} \cup \{A\})-U(P^\pi_{\{A, B\}} \cup \{B\}))
\end{split}
\]
is an unbiased estimate of $\Psi_\mathbb{O}(A,B)$.
\proof
By linearity of expectation and uniformity of samples,
\[
\begin{split}
& \mathbb{E}[\Delta \psi_{\mathbb{O}}^{A-B}]\\
=& \mathbb{E}[\frac{|\mathbb{O}|}{2|X|}\sum_{\pi \in X} \frac{1}{|\mathbb{O}|-|P^\pi_{\{A, B\}}|-1}(U(P^\pi_{\{A, B\}} \cup \{A\})\\
&-U(P^\pi_{\{A, B\}} \cup \{B\}))]\\
% =& \frac{|\mathbb{O}|}{2|X|}\mathbb{E}[\sum_{\pi \in X} \frac{1}{|\mathbb{O}|-|P^\pi_{\{A, B\}}|-1}(U(P^\pi_{\{A, B\}} \cup \{A\})\\
% &-U(P^\pi_{\{A, B\}} \cup \{B\}))]\\
=& \frac{|\mathbb{O}|}{2|X|}|X| \times \mathbb{E}[\frac{1}{|\mathbb{O}|-|P^\pi_{\{A, B\}}|-1}(U(P^\pi_{\{A, B\}} \cup \{A\})\\
&-U(P^\pi_{\{A, B\}} \cup \{B\}))]\\
=& \frac{|\mathbb{O}|}{2}\sum_{\pi \in \Pi(\mathbb{O})} \frac{1}{|\mathbb{O}|!} \times (\frac{1}{|\mathbb{O}|-|P^\pi_{\{A, B\}}|-1}(U(P^\pi_{\{A, B\}} \cup \{A\})\\
&-U(P^\pi_{\{A, B\}} \cup \{B\})))\\
=& \frac{1}{2(|\mathbb{O}|-1)!}\sum_{\pi \in \Pi(\mathbb{O})} \frac{1}{|\mathbb{O}|-|P^\pi_{\{A, B\}}|-1}(U(P^\pi_{\{A, B\}} \cup \{A\})\\
&-U(P^\pi_{\{A, B\}} \cup \{B\})) \\
=& \Psi_\mathbb{O}(A, B)\qed
\end{split}
\]
\end{corollary}

Using Corollary~\ref{cor:diff-est} in Algorithm~\ref{algo:bruteforce}, we can obtain a Monte Carlo baseline algorithm for the counterfactual explanation problem.  The pseudo-code is given in Algorithm~\ref{algo:mc-baseline}.

\begin{algorithm}[t]
\DontPrintSemicolon
%\setstretch{1.3}
\SetAlgoLined
\KwInput{the same as Algorithm~\ref{algo:bruteforce}
  }
\KwOutput{An approximation to solution to Equation~\ref{eq:problem}}
\For{$i=1$ \textbf{\em to} $|A|-1$}{
    \For {$\Delta A \subset A$ s.t. $|\Delta A|=i$}{
    Let $\mathbb{O'}= \mathbb{O}\setminus \{A, B\} \cup \{A \setminus \Delta A, B \cup \Delta A\}$;\\
    Estimate $\Delta \psi_{\mathbb{O'}}^{A \setminus \Delta A-B \cup \Delta A}$ as an unbiased estimate of $\Psi_\mathbb{O'}(A \setminus \Delta A, B \cup \Delta A)$ using Monte Carlo based on Corollary~\ref{cor:diff-est}\\
    \If{$\Delta \psi_{\mathbb{O'}}^{A \setminus \Delta A-B \cup \Delta A} < 0$}
    {
       \Return{$\Delta A$
    }
  }
}
}
\Return{A}
\caption{The Monte Carlo Baseline Algorithm \label{algo:mc-baseline}}
\end{algorithm}

\subsection{A Greedy Approach: the Framework}\label{sec:greedy}

The Monte Carlo baseline algorithm (Algorithm~\ref{algo:mc-baseline}) still needs to search across many subsets of the data sets in ascending order and thus needs to use Monte Carlo estimation many times. 
\nop{For example, if data owner $A$ owns 100 data entries and  the counterfactual explanation has 10 entries, the Monte Carlo baseline algorithm has to try out at least \re{$\sum_{i=1}^{9}{100 \choose i}+1>1.9 \times 10^{12}$ subsets of $A$}. In general, Algorithm~\ref{algo:mc-baseline} has to check \re{$\sum_{i=1}^{|\Delta A|-1}{|A| \choose {|\Delta A|-1}}+1$ subsets of $A$}, where $|\Delta A|$ is the \re{size of the} minimal explanation found, which is at least as large as the true counterfactual explanation.}

To tackle the computational cost, we explore a greedy approach. We iteratively identify the best data entry owned by data owner $A$ such that, if moved to $B$, reduces the difference of the Shapley values between $A$ and $B$ most. As the goal is to bring down the differential Shapley $\Psi_\mathbb{O}(A, B)$ from Theorem \ref{thm:differential} to a  negative value as quickly as possible, we move these ``powerful'' data entries one by one from $A$ to $B$ until $\Psi_\mathbb{O}(A, B)<0$.

Specifically, 
%in the counterfactual explanation problem of Shapley value, where there are two data owners $A, B \in \mathbb{O}$ such that $\psi_\mathbb{O}(A)>\psi_\mathbb{O}(B)$, 
for a data entry $x \in A$, the change of the differential Shapley value caused by moving $x$ from $A$ to $B$ is 
\begin{equation}\label{eq:x-diff}
\begin{split}
\Lambda_\mathbb{O}^{A \xrightarrow{x} B}=& \Psi_\mathbb{O}(A, B)-\Psi_\mathbb{O'}(A \setminus \{x\}, B \cup \{x\})\\
=&[\psi_{\mathbb{O}}(A)-\psi_{\mathbb{O}}(B)]-
[\psi_{\mathbb{O}'}(A \setminus \{x\})-\psi_{\mathbb{O}'}(B \cup \{x\})],
\end{split}
\end{equation}
where $\mathbb{O}'=\mathbb{O}[A\xrightarrow{\{x\}}B]$.
The larger the value of $\Lambda_\mathbb{O}^{A \xrightarrow{x} B}$, the more significant the change for counterfactual explanation.
%Hereafter, for the sake of brevity, we often write $\mathbb{O}[A\xrightarrow{\{x\}}B]$ as $\mathbb{O}[A\xrightarrow{x}B]$ by omitting the set parentheses of a set with single item when there is no ambiguity from context.  

Given data owners $A$ and $B$, $\Psi_{\mathbb{O}}(A, B)$ in Equation~\ref{eq:x-diff} is a constant.  Thus, we define the \textbf{power} of $x$ with respect to $A$ and $B$, denoted by $power^{\mathbb{O}}_{A\rightarrow B}(x)$, as the change of the differential Shapley value when $x$ is moved from $A$ to $B$, that is, 
\[
power^{\mathbb{O}}_{A\rightarrow B}(x)=\psi_{\mathbb{O}'}(B \cup \{x\})-\psi_{\mathbb{O}'}(A \setminus \{x\})
\]
Heuristically, the larger the power of a data entry, the more the data entry contributes to a counterfactual explanation of the Shapley values.

The framework of our greedy approach works as follows. We find the data entry $x$ that has the largest power and move it from $A$ to $B$.  If the Shapley value of $B$ becomes larger than that of $A$ after the move, that is, 
$\psi_{\mathbb{O}'}(A \setminus \{x\})<\psi_{\mathbb{O}'}(B \cup \{x\})$ and thus
$power^{\mathbb{O}}_{A\rightarrow B}(x)>0$, then $x$ is a counterfactual explanation. If not, then we iteratively find the next entry with the largest power, append this entry to the set containing the previously computed best entries, and conduct the move.  The iteration continues until the Shapley value of $B$ becomes larger than that of $A$ after the moves.  The set of all entries moved form a greedy approximation of the counterfactual explanation.

\subsection{Computing the Power of a Data Entry}

%\todo{Reviewer 2 says math is difficult to read, needs more text and less equations.}

In our greedy approach, the key operation that is performed again and again is computing the power of a data entry. Now, let us consider how to compute the power of a data entry efficiently.  

First of all, if data owner $A$ has only one data entry, that is, $A=\{x\}$, then after moving $x$ to $B$, $A$ becomes empty and thus the Shapley value is $0$.  $x$ is the trivial counterfactual explanation.  Thus, in the rest of the discussion, we assume $|A| >1$.

A data entry $x \in A$ must be in one of the two cases, $x \in A \cap B$ and $x \in A \setminus B$.  We consider the two situations one by one.

\subsubsection{The power of a Common Data Entry}

Suppose $x \in A \cap B$, that is, $x$ is a \textbf{common data entry} to $A$ and $B$.  If we move $x$ from $A$ to $B$, what change would happen to the Shapley values of the two data owners?

\nop{
According to the definition, we have
\[
power^{\mathbb{O}}_{A\rightarrow B}(x) =\psi_{\mathbb{O}'}(B \cup \{x\})-\psi_{\mathbb{O}'}(A \setminus \{x\})
\]
where $\mathbb{O}'=\mathbb{O}[A\xrightarrow{\{x\}}B]$.}  
Since $x \in A \cap B \subseteq B$, $B \cup \{x\}=B$.  Thus, using Theorem~\ref{thm:differential}, we have
\begin{equation}\label{eq:temp1}
%\begin{split}
power^{\mathbb{O}}_{A\rightarrow B}(x) =
%\psi_{\mathbb{O}'}(B)-\psi_{\mathbb{O}'}(A \setminus \{x\})\\
\sum_{\mathcal{S} \subseteq \mathbb{O}'\setminus\{B, A\setminus\{x\}\}}\frac{U(\mathcal{S}\cup\{B\})-U(\mathcal{S}\cup\{A\setminus\{x\}\}}{(|\mathcal{S}|+1){{|\mathbb{O}'|-1}\choose{|\mathcal{S}|+1}}}
%\end{split}
\end{equation}

Since $|A|>1$, $A\setminus\{x\} \neq \emptyset$, moving a data entry $x$ from $A$ to $B$ does not make $A$ empty. Thus, the number of non-empty data owners remains the same after the moving, that is$,|\mathbb{O}|=|\mathbb{O}'|$. Moreover, for every coalition $\mathcal{S} \subseteq \mathbb{O}\setminus\{A, B\}$, there exists a unique coalition $\mathcal{S}' \subseteq \mathbb{O}'\setminus\{A\setminus\{x\}, B\}$ such that $\mathcal{S}=\mathcal{S}'$, and vice versa. Based on these two observations, Equation~\ref{eq:temp1} can be further rewritten to
\begin{equation}\label{eq:temp2}
\begin{split}
& power^{\mathbb{O}}_{A\rightarrow B}(x)\\
%=& \sum_{\mathcal{S} \subseteq \mathbb{O}\setminus\{B, A\setminus\{x\}\}}\frac{U(\mathcal{S}\cup\{B\})-U(\mathcal{S}\cup\{A\setminus\{x\}\}}{(|\mathcal{S}|+1){{|\mathbb{O}|-1}\choose{|\mathcal{S}|+1}}}\\
=& \sum_{\mathcal{S} \subseteq \mathbb{O}\setminus\{B, A\setminus\{x\}\}}\frac{U(\mathcal{S}\cup\{B\})}{(|\mathcal{S}|+1){{|\mathbb{O}|-1}\choose{|\mathcal{S}|+1}}}-\sum_{\mathcal{S} \subseteq \mathbb{O}\setminus\{B, A\setminus\{x\}\}}\frac{U(\mathcal{S}\cup\{A\setminus\{x\}\}}{(|\mathcal{S}|+1){{|\mathbb{O}|-1}\choose{|\mathcal{S}|+1}}}
\end{split}
\end{equation}

\nop{
The first term in Equation~\ref{eq:temp2} is constant for any $x \in A \cap B$.  Thus, we have 
\begin{equation}\label{eq:temp5}
power^{\mathbb{O}}_{A\rightarrow B}(x) = -\sum_{\mathcal{S} \subseteq \mathbb{O}\setminus\{B, A\setminus\{x\}\}}\frac{U(\mathcal{S}\cup\{A\setminus\{x\}\}}{(|\mathcal{S}|+1){{|\mathbb{O}|-1}\choose{|\mathcal{S}|+1}}}
\end{equation}
}

To enable Monte Carlo approximation, using Corollary~\ref{cor:mc-baseline}, we have:
\begin{equation}\label{eq:temp6}
\begin{split}
& power^{\mathbb{O}}_{A\rightarrow B}(x) = \psi_{\mathbb{O}'}(B \cup \{x\})-\psi_{\mathbb{O}'}(A \setminus \{x\})\\
=& \psi_{\mathbb{O}'}(B)-\psi_{\mathbb{O}'}(A \setminus \{x\})\\
=& \frac{1}{2(|\mathbb{O}'|-1)!}\sum_{\pi \in \Pi(\mathbb{O}')} \frac{1}{|\mathbb{O}'|-|P^\pi_{\{A\setminus \{x\}, B\}}|-1}\Big (U(P^\pi_{\{A\setminus \{x\}, B\}} \cup \{B\})\\
& -U(P^\pi_{\{A\setminus \{x\}, B\}} \cup \{A\setminus \{x\}\}\Big )\\
\end{split}
\end{equation}
Notice that, for any permutation $\pi' \in \Pi(\mathbb{O}')$, there exists a unique permutation $\pi \in \Pi(\mathbb{O})$ such that $P^{\pi'}_{\{A\setminus \{x\}, B\}}=P^{\pi}_{\{A, B\}}$ and vice versa. Thus, Equation~\ref{eq:temp6} can be further rewritten to
\begin{equation}\label{eq:temp7}
\begin{split}
& power^{\mathbb{O}}_{A\rightarrow B}(x) \\
=& \frac{1}{2(|\mathbb{O}|-1)!}\sum_{\pi \in \Pi(\mathbb{O})} \frac{U(P^\pi_{\{A, B\}} \cup \{B\})-U(P^\pi_{\{A, B\}} \cup \{A\setminus \{x\}\}}{|\mathbb{O}|-|P^\pi_{\{A, B\}}|-1}\\
% =&  \frac{1}{2(|\mathbb{O}|-1)!}\sum_{\pi \in \Pi(\mathbb{O})} \frac{U(P^\pi_{\{A, B\}} \cup \{B\})}{|\mathbb{O}|-|P^\pi_{\{A, B\}}|-1}\\
% & -  \frac{1}{2(|\mathbb{O}|-1)!}\sum_{\pi \in \Pi(\mathbb{O})} \frac{U(P^\pi_{\{A, B\}} \cup \{A\setminus \{x\}\}}{|\mathbb{O}|-|P^\pi_{\{A, B\}}|-1}\\
\end{split}
\end{equation}

\nop{
The first term in Equation~\ref{eq:temp7} is constant for any $x \in A \cap B$.  Thus, we have 
\begin{equation}\label{eq:temp8}
power^{\mathbb{O}}_{A\rightarrow B}(x) \sim -  \frac{1}{2(|\mathbb{O}|-1)!}\sum_{\pi \in \Pi(\mathbb{O})} \frac{U(P^\pi_{\{A, B\}} \cup \{A\setminus \{x\}\}}{|\mathbb{O}|-|P^\pi_{\{A, B\}}|-1}
\end{equation}

Equations~\ref{eq:temp5} and~\ref{eq:temp8} are intuitively understandable. Since $x \in A \cap B$, removing $x$ from $A$ reduces the contributions of $A$ to coalitions but does not improve those of $B$.  Therefore, the power of $x$ with respect to $A$ and $B$ depends only on how much removing $x$ from $A$ reduces the contributions of $A$ to coalitions.
}

Based on Corollary~\ref{cor:diff-est} and Equation~\ref{eq:temp7}, we have the following Monte Carlo estimation.

\begin{theorem}[MC-Common Entry]\label{thm:mc-common}
For $x \in A \cap B$ and a uniform sample of permutations $X \subseteq \Pi(\mathbb{O})$,
\[
\widehat{power^{\mathbb{O}}_{A\rightarrow B}(x)}=
\frac{|\mathbb{O}|}{2|X|}\sum_{\pi \in X} \frac{U(P^\pi_{\{A, B\}} \cup \{B\})-U(P^\pi_{\{A, B\}} \cup \{A\setminus \{x\}\}}{|\mathbb{O}|-|P^\pi_{\{A, B\}}|-1}
\]
is an unbiased estimation of $power^{\mathbb{O}}_{A\rightarrow B}(x)$.
\qed
\end{theorem}

\subsubsection{The power of a Differential Data Entry}

Suppose $x \in A \setminus B$, that is, $x$ is a data entry that $A$ has but $B$ does not.  In such a case, we call $x$ a \textbf{differential data entry}. Moving $x$ from $A$ to $B$ may not only reduce the contributions from $A$ to coalitions but may also improve those from $B$. Using the exact same logic and theorems as above, we get the following Monte Carlo estimation:

\begin{theorem}[MC-Differential Item]\label{thm:mc-diff}
For $x \in A \setminus B$ and a uniform sample of permutations $X \subseteq \Pi(\mathbb{O})$,
\[
\begin{split}
\widehat{power^{\mathbb{O}}_{A\rightarrow B}(x)} =&  \frac{|\mathbb{O}|}{2|X|}\sum_{\pi \in X} \frac{1}{|\mathbb{O}|-|P^\pi_{\{A, B\}}|-1} \\
& \cdot \Big (U(P^\pi_{\{A, B\}} \cup \{B\cup\{x\}\})-U(P^\pi_{\{A, B\}} \cup \{A\setminus \{x\}\})\Big )\\
\end{split}
\]
is an unbiased estimation of $power^{\mathbb{O}}_{A\rightarrow B}(x)$.
\qed
\end{theorem}

\subsection{The SV-Exp Algorithm}

After carefully assembling all the necessary components, we are now ready to introduce our greedy algorithm for a counterfactual Shapley value explanation, SV-Exp.  

\subsubsection{Framework}

SV-Exp works in two phases.  In the first phase, we conduct Monte Carlo estimation to find the top-1 data entry in $A$ that has the largest power.  Then, in the second phase, we move the top-1 data entry from $A$ to $B$.  After moving the top-1 data entry from $A$ to $B$, if $\psi(A) > \psi(B)$ still holds, we repeat the process, that is, finding the next top data entry and moving it from $A$ to $B$, until the Shapley value relationship is reversed.  The pseudocode is shown in Algorithm~\ref{algo:sv-exp}.

\begin{algorithm}[t]
\DontPrintSemicolon
%\setstretch{1.3}
\SetAlgoLined
\KwInput{the same as Algorithm~\ref{algo:bruteforce} and, in addition, a confidence threshold $0 < \delta < 1$, and a threshold for the top-1 data entry's confidence interval $\epsilon$
  }
\KwOutput{An approximation solution to Equation~\ref{eq:problem}}
Counterfactual explanation $\Delta A \gets \emptyset$;\\
Estimate $d=\psi_\mathbb{O'}(A \setminus \Delta A) - \psi_\mathbb{O'}(B \cup \Delta A)$ using Monte Carlo based on Corollary~\ref{cor:diff-est}\\
\While{$d$ is not converged and $d>0$}{
  // Phase 1: finding the best data entry in $A$ \\
  \For{$x \in A$}{
     $power_x \gets 0$ and $\delta$-confidence interval $interval_x \gets [0, \infty]$;\\
     }
  \While{the $\delta$-confidence interval of the top-1 entry in $A$ has size larger than $\epsilon$}{
    use Thompson sampling to find the approximate best data entry $x_{best}$, let $p$ be the estimated probability that $x_{best}$ is indeed the best entry \\
    $x \gets x_{best}$ with probability $p$ and a random data entry in $A$ other than $x_{best}$ with probability $(1-p)$\\
    draw a uniform sample of permutations from $\Pi(\mathbb{O})$; \\
      estimate and update $power^{\mathbb{O}}_{A \rightarrow B}(x)$ and the $\delta$-confidence interval using Theorems~\ref{thm:mc-common} and~\ref{thm:mc-diff};\\
   }
   $x \gets \text{top-1 data entry in A}$\\
   $\;$
   // Phase 2: producing counterfactual explanation in a greedy manner \\
   $\Delta A \gets \Delta A \cup \{x\}$\\
     $\mathbb{O} \gets (\mathbb{O}\setminus \{A, B\}) \cup \{A \setminus \{x\}, B\cup \{x\}\}$ ;\\
     $A \gets A \setminus \{x\}$, $B \gets B \cup \{x\}$\\
%     $d \gets$ Monte Carlo estimate\\
        estimate $d=\psi_\mathbb{O'}(A \setminus \Delta A) - \psi_\mathbb{O'}(B \cup \Delta A)$ using Monte Carlo based on Corollary~\ref{cor:diff-est}\\
      \If{$d$ converged and $d< 0$}{\Return{$\Delta A$}}
      %\Else{remove the top-ranked data entry $x$ from $A$}
}
\caption{The SV-Exp Algorithm \label{algo:sv-exp}}
\end{algorithm}

\subsubsection{Phase 1}

In Phase 1, SV-Exp iteratively draws a uniform sample of permutations and then uses the Monte Carlo approach to estimate the power value of a selected data entry in $A$ so that the $\delta$-confidence interval is no more than $\epsilon$ (e.g., $\delta = 95\%$ and $\epsilon=0.01$).

We use \textbf{Thompson sampling}~\cite{10.1093/biomet/25.3-4.285, af40c05f-9359-3ac3-8ebd-65d7de239b60, russo2020tutorial} to accommodate the explore-exploit nature of the problem. The explore part of the problem arises from the fact that we must sample and estimate the powers of different data entries to find the ``true'' best data entry. The exploit part of the problem arises from our desire to actually confirm that the data entry currently ranked first is \emph{indeed} the best, which we can only find out by continuing to sample for that data entry and becoming more confident about its power. Thus, there is a trade-off between sampling heavily for our current-ranked-first data entry and sampling widely for other data entries in $A$'s data in the case another entry happens to be a better top-1 item. Thompson sampling addresses both these concerns, making it an effective algorithmic choice for this problem.

Thus, we begin by sampling a small amount of permutations and estimating the differential Shapley value for each entry to get our prior. Then, we sort the data entries by power and pick a data entry for further sampling. We pick this data entry by approximating a normal distribution over the power of each data entry using the current means and standard deviations of each draw, drawing one random power value from each distribution, sorting the randomly drawn power values (one associated with each data entry $x \in A$), and selecting the data entry associated with the best power.

For the selected data entry, we draw a uniform batch of samples of its powers using Theorems \ref{thm:mc-common} and \ref{thm:mc-diff}, update the differential Shapley value, sort, and check if the power of the first-place data entry has $\delta$-confidence interval no more than $\epsilon$.
%and that of the second-place data entry} are significantly different at a \re{$\delta=95\%$} significance level. If so, we are $\delta$-confident that our top-1 entry is indeed the most powerful. 
This Bayesian approach to ranking the data entries is efficient due to our objective of finding the top-1 data entry in every round. This means that we are unlikely to have to fully estimate the power for every single data entry. This is a huge advantage of Thompson sampling and is extremely time efficient compared to the frequentist approach, which would require sampling the powers of every single data entry the same amount of times until the first-place and second-place data entries had significantly different powers. With Thompson sampling, we utilize our posterior beliefs about the best data entry every single time we sample. 

At the end of Phase 1, we have our estimated best data entry to shift.

% The estimation of power uses Theorems~\ref{thm:mc-common} and~\ref{thm:mc-diff}.  
% Moreover, 
% The first set of permutations in the sample is used to estimate the power of every data entry in $A$, which is our prior. In Thompson sampling, the hope is that over time we will only continue to draw samples for the top few data entries until we are $\delta$-confident in their true differences with one another. This allows us to stop sampling the less powerful data entries early in the process to improve efficiency.

\subsubsection{Phase 2}

In this phase, we move the top-1 data entry from $A$ to $B$.  After the shift, we check whether the Shapley value relationship is reversed.  We use Corollary~\ref{cor:diff-est} to check whether with $\delta$-confidence, $\psi(A') < \psi(B')$ holds.  If so, we have successfully obtained a counterfactual explanation, and the algorithm terminates.  Otherwise, we repeat phase 1.
%We add this next data entry to our subset \re{of counterfactual explanation} in consideration., which will now be (first best data entry, second best data entry) and check if the Shapley value relation has flipped, 
We continue this iteration between Phase 1 and Phase 2 until a valid counterfactual is found.

\section{Experiments}\label{sec:exp}

In this section, we report experimental results on three real datasets to examine the effectiveness of the counterfactual explanation of Shapley values and the efficiency of our method.

\subsection{Experimental Setup}
The experiments were run on the Duke Computing Cluster (DCC) using Slurm. We used nodes from the 10x TensorEX TS2-673917-DPN Intel Xeon Gold 6226 Processor, 2.7Ghz (768GB RAM 48 cores). Each of these machines has 2 Nvidia GeForce 2080 RTX Tis.

%\subsubsection{Methods}
We implemented and compare two methods, the Monte Carlo baseline (Algorithm~\ref{algo:mc-baseline}, denoted by MC) and our SV-Exp algorithm (Algorithm~\ref{algo:sv-exp}). Both were implemented in Python. 

%\subsubsection{Datasets} 
We utilized three different datasets. The Breast Cancer Wisconsin Dataset from the UCI Machine Learning Repository~\cite{misc_breast_cancer_14} has 455 records for training and 63 records for testing, both in 32 attributes. The Boston Housing Prices dataset accessed through Kaggle~\cite{HARRISON197881} has 354 records for training and 152 records for testing, both in 14 attributes. The Hotel Reservations Dataset, also accessed through Kaggle~\cite{ANTONIO201941}, has 800 records for training and 200 records for testing in 19 attributes. This 1,000-record dataset was a uniform sample without replacement of the original 36,275 records, which was then split using the train-test-split function from Sklearn (\url{https://scikit-learn.org}). 
%Since the Hotel Reservations Dataset is our example of a real-world dataset and its main role in the paper is to have grounded, interpretable results, 
% The reason to obtain the uniform samples of the Hotel Reservations Dataset is simply to scale down the dataset and data owner sizes to make the results more approachable and understand. Our method can be applied to the entire original dataset.
 
% For each trial, a sampled dataset $D$ is used, and a certain number of data owners are set up. Each data owner is assigned a uniform sample of $D$. We randomly choose two data owners $A$ and $B$ such that $\phi(A)>\phi(B)$. 

%For the Zipfian and Natural Case Study scenarios, 10 trials were run for each assignment due to the higher level of consistency of assignment, sizes of A and B, and coalition structure. 

%\subsubsection{Utility Function}
The utility of a set of data entries is as follows: given some task (kernel density estimation, logistic regression, etc.), we use the set of data entries as a training set for the specified task. The utility is exactly the measured performance of the task on the test set. We used four different tasks and the associated utility functions in our experiments: kernel density estimation ($\eta-$sum-of-absolute errors), logistic regression prediction ($\eta-$log loss), random forest regression ($\eta-$MSE), and linear regression ($\eta-$MSE), where $\eta$ is a sufficiently large number so that the utility value is non-negative.

\nop{
\begin{algorithm}[t]
\DontPrintSemicolon
%\setstretch{1.3}
\SetAlgoLined
\KwInput{Data Owner $O$, training data $TR$, test data $TS$, utility function $k$
}
\KwOutput{Prediction performance of the model trained on the data owner's}
Let owner data $S$ = subset of $TR$ owned by $O$. \\
Train model $M$ using utility function $k$ on $S$, $pred =$ predicted labels of $M$ on $TS$\\
\Return{-MSE, -Log. Loss, or Accuracy ($\frac{\text{correctly predicted}}{\text{correctly predicted}}$)depending prediction task}
\caption{Utility Algorithm \label{algo:util}}
\end{algorithm}
}

\subsection{Efficiency}

%In this section, we investigate the efficiency of the two methods on the real datasets under various settings.

\begin{figure}[t]
    \centering
      \begin{subfigure}
        {0.49\linewidth}
        \includegraphics[width=\linewidth]{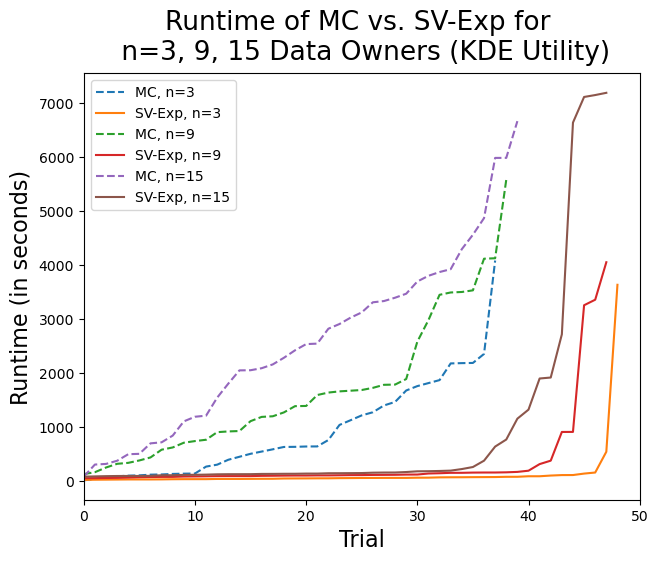}
          \caption{KDE}
          \label{fig:runtimeplotkde}
      \end{subfigure}
      \hfill
      \begin{subfigure}
        {0.49\linewidth}
        \includegraphics[width=\linewidth]{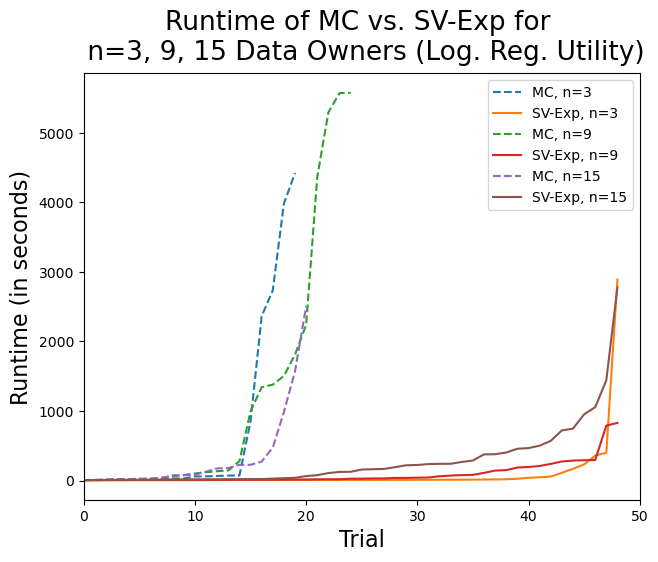}
          \caption{Logistic Regression}
          \label{fig:runtimeplotlr}
      \end{subfigure}
\caption{Runtime for $n=3,9,15$ data owners for different utility functions. For each method, trials were sorted in runtime ascending order.}
\label{fig:runtimecomp}
\end{figure}

%\subsubsection{Uniform Distribution}
We test the runtime of the MC and the SV-Exp algorithms using the Breast Cancer Wisconsin dataset. We set the number $n$ of data owners to 3, 6, 9, 12, and 15, respectively. For each set of parameters in the Breast Cancer dataset, 50 trials were run. We use KDE and logistic regression as our two utility functions, as KDE is a very local function relative to logistic regression. We diversify our utility functions to show the generalizability of our methods.

Each data owner's data was drawn uniformly without replacement from the whole training set. The size of each data owner's dataset is distributed uniformly over the range $[1, 455]$. We allow each run to take up to 7,200 seconds. If a method in a test cannot finish in 7,200 seconds, then the trail is marked as timeout.

Limited by space, we only show the results of $n=3, 9, 15$ of KDE and logistic regression in Figure~\ref{fig:runtimecomp}. SV-Exp is much faster than MC for every set of parameters, as MC times out in many trials.

%Let us analyze the performance of the algorithms over two utility functions. 
The general trend is that the runtime increases as the number of data owners $n$ increases.
%The performance of the two counterfactual approximation algorithms is very dependent on the number of data owners. 
When there are many data owners with uniformly distributed size and uniformly distributed data entries, each owner's marginal contribution lessens as the number of data owners increases. Thus, we expect counterfactual sizes to be small in the presence of many data owners. When there are a small number of data owners, each owner's marginal contribution is highly correlated with the size of their dataset, so we expect a large counterfactual between owners.
%whose dataset ownership proportion is large. 
% Thus, SV-Exp naturally performs better since MC is expected to test many subsets of increasing size, increasing runtime exponentially. 

% The KDE utility function has good locality, while logistic regression is less dependent on the locality of data. Consequently, changing the number of data owners while using logistic regression for utility doesn't impact the runtimes between MC and SV-Exp very much.  Therefore, in much more trials MC times out. SV-Exp outperforms MC with a much larger advantage than in the logistic regression case.
%, with MC timing out on more than half the trials and SV-Exp not only finishing the task but with much lower runtime. %\todo{why might the kde have more of a linear trend but the log reg seems extremely sharp and exponential for the monte carlo algorithm? couldn't figure out how to explain this}

Figure~\ref{fig:barplots} shows the average runtime with respect to the number of data owners in all 5 settings. SV-Exp consistently outperforms MC in mean runtime. The standard deviations of both methods are big, as expected, due to the exponential nature of the problem and the randomness in how data owners $A$ and $B$ are chosen and the sampling of data coalitions. 

% \begin{figure}

\begin{figure}[t]
    \centering
      \begin{subfigure}
        {0.49\linewidth}
        \includegraphics[width=\linewidth]{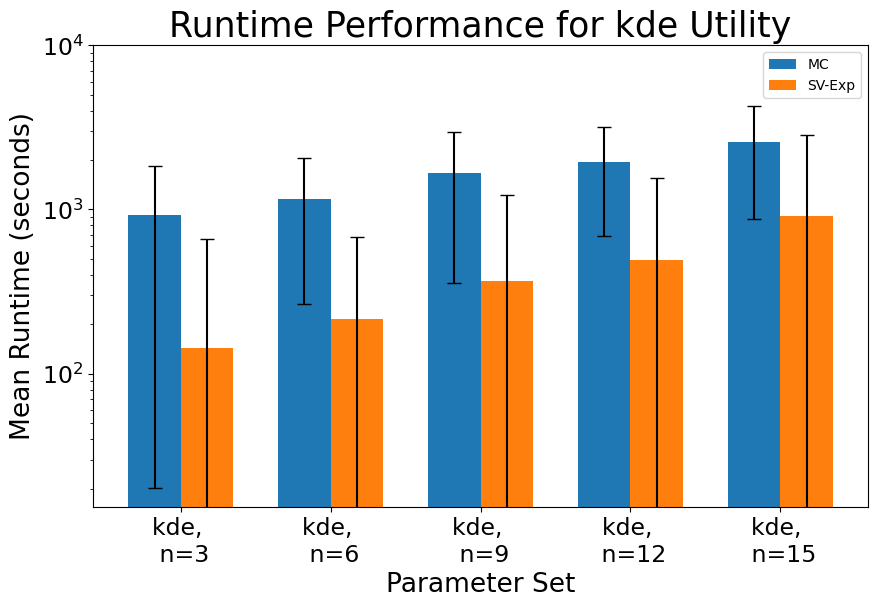}
          \caption{KDE}
          \label{fig:barplotkde}
      \end{subfigure}
      \hfill
      \begin{subfigure}
        {0.49\linewidth}
        \includegraphics[width=\linewidth]{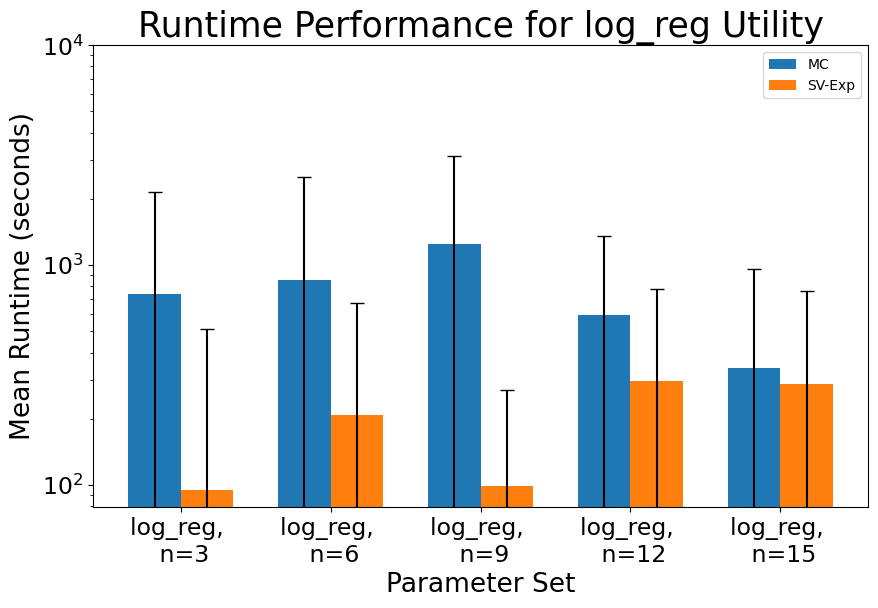}
          \caption{Logistic Regression}
          \label{fig:barplotlr}
      \end{subfigure}
\caption{Runtime statistics, where the Y-axis is in logarithmic scale.}
\label{fig:barplots}
\end{figure}

\begin{figure}[t]
    \centering
\nop{      \begin{subfigure}
        {0.49\linewidth}
        \includegraphics[width=\linewidth]{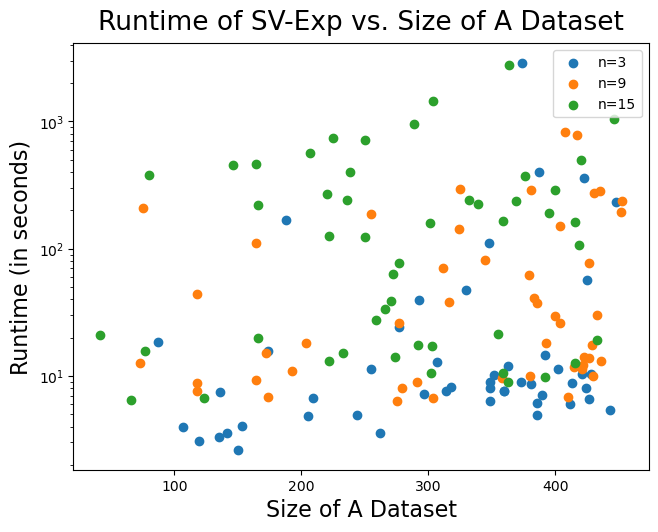}
          \caption{Size of $A$}
          \label{fig:asizeruntime}
      \end{subfigure}
      \hfill
      \begin{subfigure}
        {0.49\linewidth}
        \includegraphics[width=\linewidth]{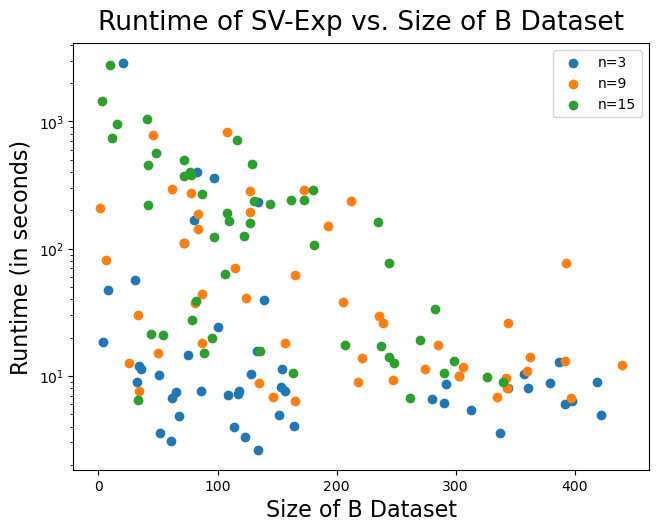}
          \caption{Size of $B$}
          \label{fig:bsizeruntime}
      \end{subfigure}
}        \begin{subfigure}
        {0.49\linewidth}
        \includegraphics[width=\linewidth]{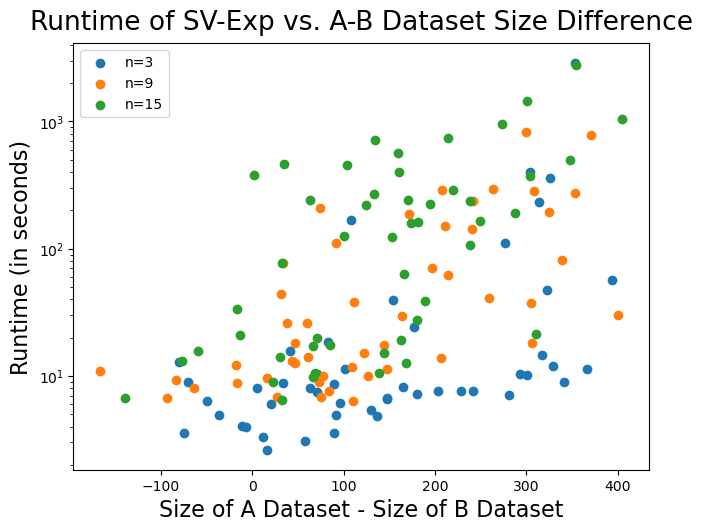}
          \caption{Size of $|A| -|B|$}
          \label{fig:abdiffruntime}
      \end{subfigure}
      \begin{subfigure}
        {0.49\linewidth}
        \includegraphics[width=\linewidth]{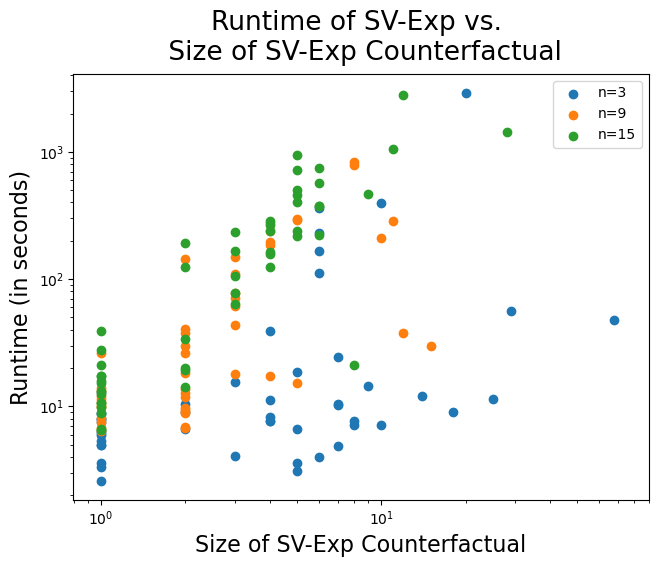}
          \caption{Size of counterfactual exp}
          \label{fig:answerruntime}
      \end{subfigure}
\caption{Runtime with respect to %sizes of data owners $A$ and $B$, 
size difference between $A$ and $B$, and size of counterfactual. Only the results of SV-Exp were shown since the majority of MC trials timed out.}
\label{fig:inputsvsruntime}
\end{figure}

Figure~\ref{fig:inputsvsruntime} shows the runtime with respect to %the sizes of data owners $A$ and $B$, 
the size difference between $A$ and $B$, and the size of counterfactual. Only the results of SV-Exp were shown since the majority of MC trials timed out.
%
%Now we analyze the runtimes in relation to inputs. 
%In Figure~\ref{fig:asizeruntime}, we observe a very slightly positive relationship between the runtime and the size of $A$. When $A$ is bigger, it is more likely to have a relatively high Shapley compared to $B$, and it takes longer to find a counterfactual subset. In this vein, Figure~\ref{fig:bsizeruntime} shows a downward trend—when $B$ is smaller, it is more likely to differ significantly in the Shapley value to $A$, so we expect to have to shift more data entries and thus it takes longer to find the counterfactual. This relationship is verified in 
Figure~\ref{fig:abdiffruntime} %, which 
clearly shows that the runtime is roughly positively correlated to the size difference between $A$ and $B$.
%size is what correlates most with difference in runtimes. 
The bigger the difference, likely the longer the runtime. Note that $B$ may have more data entries than $A$ and still may have a lower Shapley value (negative x-values), but these situations are often flipped very easily because $A$ may own one or two very powerful data entries that are driving its higher Shapley value. This is a perfect example of the motivation for finding a counterfactual explanation, which can help us glean scenarios where there are ``star player'' data entries. 
%Finally,
Figure~\ref{fig:answerruntime} demonstrates the power of the SV-Exp algorithm. Smaller counterfactuals take shorter runtime. The SV-Exp algorithm has no problem in handling counterfactuals as large as 90 data entries within the timeout limit 7,200 seconds, even with a large number of data owners.
% \vspace{-5mm}

\subsection{Effect of Number of Data Owners}

%\re{In this section, we investigate an interesting question:} 
How does the size of the counterfactual explanation change as the number of data owners increases?
%the parameters change? 
Since MC times out in many trials, we only report the results achieved by SV-Exp.

%\subsubsection{Uniform Distribution}

\begin{table}[t]
  \centering
  \caption{The average size of counterfactual explanations with respect to number of data owners, when data entries are assigned to data owners randomly in uniform distribution. Two different utility functions KDE and Logistic Regression (LG) are used. Standard deviations are reported in parentheses.} %Items are bolded where none of the trials timed out.}
  \centering
  \begin{tabular}{lcc}
    \toprule
    \textbf{$n$} & \textbf{SV-Exp-KDE} & \textbf{SV-Exp-LG} \\
    \midrule
    3  & 2.63 (6.31) & 7.02 (10.80)\\
    6  & 2.92 (6.48) & 5.40 (8.83)\\
    9  & 2.37 (4.34) & 5.22 (13.31)\\
    12 & 1.93 (2.09) & 3.55 (2.73) \\
    15 & 1.68 (1.74) & 4.06 (4.38) \\
    \bottomrule
  \end{tabular}
  \label{tab:unicounterfactual}
\end{table}

Table~\ref{tab:unicounterfactual} shows the average size of the counterfactual explanation with respect to the number of data owners in uniformly distributed data. 
%The counterfactual explanations resulting from parameter regimes in which none of the trials timed out are bolded to highlight which counterfactuals are actually interpretable. The SV-Exp counterfactuals in Table~\ref{tab: uniformcounterfactuallr} support the logic that 
As the number of data owners increases, each owner's data most likely becomes less critical, and thus the counterfactual explanation between two data owners will contains less data entries on average. 
%When $n=3$, any large difference in size between two data owners will result in a large counterfactual explanation. 
Because each data owner's size is uniformly redrawn for each trial, the standard deviation of the counterfactual explanation is large especially when $n$ is small. Note that the standard deviation also tends to decrease as $n$ increases—the more data owners, the higher the probability that a pair of randomly chosen owners have similar Shapley values.

\subsection{Effect of Allocation Among Data Owners}

To examine the effect of various allocations among data owners, we use the Zipfian distribution, where the size of each data owner's dataset is drawn from this distribution. This setting simulates real-world scenarios where a few large data owners hold most of the data, while many smaller players have much less.

%a scenario in some real world applications, where data ownership is mainly conglomerated in the hands of a few large data owners, and there are many weaker players with a much smaller share of the data. 

\begin{figure}[t]
  \centering
  \includegraphics[width=1.0\linewidth]{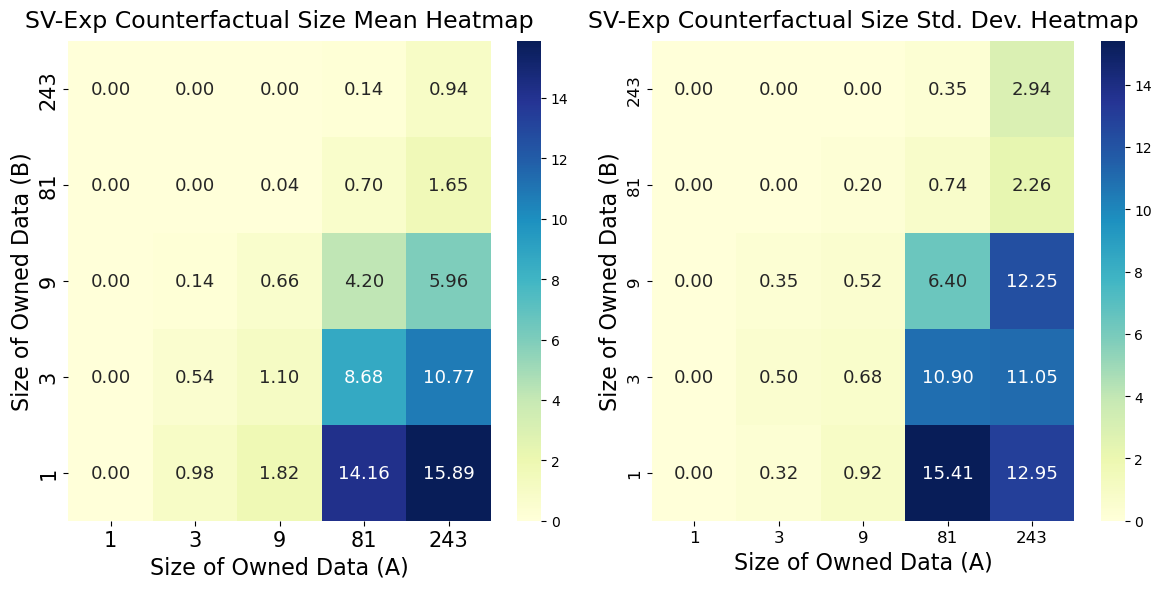}
%          \caption{SV-Exp}
%          \label{fig:zipfianlrgreedyanswer}
%      \end{subfigure}
\caption{Size of counterfactual explanations when data entries are assigned to data owners following the Zipfian Distribution. In each subfigure, the left shows the mean and the right shows the standard deviations.}
\label{fig:zipfiananswers}
\end{figure}

We report the results on two data owners in experiments, one with a dataset size $n_1$ such that $log_a n_1 = k_1$ and another with a size $n_2$ such that $\log_a n_2 = k_2$. To match the scale of the Breast Cancer Wisconsin training dataset, we set $a=3$ and let $k \in \{0,\dots,4\}$. We fix the number of data owners to 9 (the median value from the uniform distribution experiments) and test the 25 possible pairings of data owner sizes for all values of $k$ over 50 trials. The results are shown in Figure~\ref{fig:zipfiananswers}. %Trials where the algorithm had any case of timing out were left blank. %, as we have no interpretation for a counterfactual for these cases. 
Note that the left upper triangle of the matrices have counterfactual explanations of size 0 because $A$ and $B$ fail to satisfy the precondition that $\psi(A) > \psi(B)$.
The counterfactual explanations of the largest sizes happen when $A$ owns a large portion of the data and $B$ owns a small portion. We see from Figure~\ref{fig:zipfiananswers} that in a case where the whole data set has 455 training records and 63 testing records, the counterfactual explanation computed by SV-Exp between a data owner $A$ who owns more than half of the dataset and a data owner $B$ who owns only one entry comes down to almost 16 data entries on average. Those 16 estimated ``best'' data entries provide insight on the data records that are contributing most to the difference in the predictive power of data owner $A$ versus data owner $B$. Similarly to the cases in the uniform distribution case, the standard deviation of the size of counterfactual explanation increases as the size of the counterfactual explanation increases, demonstrating the uncertainty in the difference of Shapley values at high levels of disparity between data owners.

We observe an interesting case where the size of counterfactual explanation between a data owner $A$ having 81 data entries against another data owner $B$ having only one data entry is larger than that in another case where $|A|=243$ and $|B|=3$. This speaks to some notion of decreasing returns in the Shapley value (3 items has a lot more power than 1 item, but 243 items does not have much more power than 81). Please note that the size of counterfactual explanations is very sensitive to the detailed data assignments.

\subsection{Accuracy in Finding Counterfactuals}

We use a small random subset of the Breast Cancer dataset to compare the counterfactuals given by the Brute-Force algorithm (Algorithm \ref{algo:bruteforce}) with exact Shapley value computation, which are indeed the ground truth. We compare with the results from the two approximation algorithms, MC (Algorithm \ref{algo:mc-baseline}) and SV-Exp (Algorithm \ref{algo:sv-exp}). To allow the Brute Force algorithm to finish, we extract a uniform random sample of size 40 as the training data set and another disjoint uniform random sample of size 10 as the test set. We run the experiments with 3 data owners. The data owners are assigned the data from a uniform distribution, two utility functions (KDE and Logistic Regression) were used, and 50 trials were run for each utility function. The results are shown in Tables~\ref{tab:cflength}
and \ref{tab:jaccindex}.

As expected, when dataset sizes are tiny, MC approximates both a similar size and similar data records as the Brute Force algorithm because it evaluates subsets of data owner A in the same order as the brute force method (size-ascending). 
MC even tends to underestimate the size (finishes early) compared to SV-Exp, as there is some level of error when checking if the differential Shapley is negative. Due to SV-Exp's greedy nature, it is also expected that it will overestimate the sizes of counterfactuals.%, which is consistent with both utility functions.

\nop{
The Jaccard similarity gives information about the similarity of the contents of the counterfactual in addition to the size. A Jaccard similarity index between two lists takes a value between 0 and 1, with 0 meaning no similarity and 1 meaning absolute similarity (they are the same list). The average Jaccard similarity between the BF and MC counterfactuals is much higher than the SV-Exp approximations. This is again due to the fact that MC evaluates the same subsets and in the same order as BF, whereas the nature of the SV-Exp algorithm means that data entries are evaluated individually rather than in subsets.}
%The average counterfactual length and Jaccard similarity are not agnostic to utility—using Logistic Regression as the utility function led to more similarity between the counterfactuals resulting from the two approximation algorithms versus the exact algorithm than using KDE.
The average counterfactual length and Jaccard similarity are not agnostic to utility. Using Logistic Regression as the utility function leads to more similarity between the counterfactuals from the two approximation algorithms compared to the exact algorithm than using KDE.

The results show that on tiny datasets with a small number of data owners, MC provides a better approximation of the ground-truth. However, when the dataset size and the number of data owners increase, MC quickly loses the edge due to its weak scalability. Moreover, MC tends to underestimate the size of counterfactuals. 
%Thus, our method performs better on larger datasets, and 
In the context of obtaining data in data markets, a slight overestimation of the counterfactual size is most likely more beneficial than an underestimation—-if flipping the Shapley value is the goal, it is better to err on the side of being sure that the amount of data bought will successfully flip the Shapley value than to not buy enough data and still lose to the opposing data owner.
% \begin{figure*}[t]
%   \centering
%   \includegraphics[width=0.6\linewidth]{jaccard_index_zipf.png}
%     \caption{Jaccard index of MC and SV-Exp CF answers for trials where both algorithms finished running.}
%     \label{fig:jaccard}
% \end{figure*}

\begin{table}
\centering
\caption{Average counterfactual lengths across the three algorithms: BF (Brute-Force), MC (Monte-Carlo), SV-Exp. Standard deviations are given in parentheses.}
  \begin{tabular}{lccc}
    \toprule
     \textbf{Utility Function}& \textbf{BF} & \textbf{MC} &\textbf{SV-Exp} \\
    \midrule
    KDE  & 2.50 (1.73) & 2.14 (1.19) & 4.20 (4.68)\\
    Log. Reg.  & 2.83 (1.62) & 2.60 (1.30) & 3.36 (2.68) \\
    \bottomrule
  \end{tabular}
  \label{tab:cflength}
\end{table}
\begin{table}
\centering
\caption{Average Jaccard similarity indices of counterfactuals between BF (Brute-Force) and MC (Monte-Carlo) as well as BF and SV-Exp. Standard deviations are given in parentheses.}
  \begin{tabular}{lcc}
    \toprule
     \textbf{Utility Function}& \textbf{(BF, MC)} & \textbf{(BF, SV-Exp)}\\
    \midrule
    KDE  & 0.89 (0.18) & 0.32 (0.20) \\
    Log. Reg.  & 0.92 (0.13) & 0.41 (0.18) \\
    \bottomrule
  \end{tabular}
  \label{tab:jaccindex}
\end{table}

%\subsection{Success Comparison}

%MC is a Monte Carlo approximation method to approach counterfactual explanations directly, and SV-Exp greedily searches for counterfactual explanations that may not be minimal. Now that we have compared the approximate counterfactuals against the brute-force exact counterfactuals in small examples, it is also important to
Let us now compare using datasets whose sizes are not tiny whether the approximated counterfactuals by MC and SV-Exp actually reverse the Shapley value relation successfully.
%
% We do observe that the top data entries identified by SV-Exp always appear in the answers produced by MC. At the same time, please note that, since on those datasets we do not have any ground-truth, comparing the data entries in the answers generated by the two methods\footnote{We have these results but omit them here due to space limitations.} does not provide a correct picture due to the nature of those methods being approximation approaches. Instead, we compare how well those methods provide counterfactual explanations that can invert the inequality between the Shapley values of two data owners.
%
Specifically, after each trial, $\psi(A \setminus {\Delta A})-\psi(B \cup \Delta A)$ was estimated using Monte Carlo, where $\Delta A$ is the answer generated by MC or SV-Exp. If the difference was negative, the trial was marked a success and a failure otherwise. Both MC and SV-Exp were set to time out at 7,200 seconds. Those that timed out were not counted in the comparison: only successful trials were used in calculating the average success rate of the methods so as to not doubly penalize MC for timeouts. 
Note that our measure for ``success'' is also an approximation, as we cannot check the exact Shapley value difference when the datasets are not tiny. However, checking the approximate differential Shapley value after the approximate $\Delta A$'s are produced still gives a good sense about whether our resulting counterfactuals reverse the Shapley value relation, since the Monte Carlo estimation is the state-of-the-art approach in practice. 
%as any false negative means that a significant amount of coalitions still lead to a negative Shapley.

%\subsubsection{Uniform Distribution}
%\todo{again, not sure if we need this or if it can be made smaller since i know we're out of space}

\begin{table}[t]
  \centering
  \caption{Success rates of finding counterfactual explanations under uniformly distributed data owners in KDE and Logistic Regression (LG) tasks. Standard deviations are given in parentheses.}\label{tab:accuracy}
%\small
%  \begin{subtable}[]{0.45\textwidth}
  \centering
  \begin{tabular}{l|cc|cc}
    \toprule
    \textbf{$n$} & \textbf{MC-KDE} & \textbf{SV-Exp-KDE} & \textbf{MC-LG} & \textbf{SV-Exp-LG}\\
    \midrule
    3  & 0.95 (0.23) & 0.98 (0.14) & 0.90 (0.31) & 0.88 (0.33)  \\
    6  & 0.98 (0.16) & 1.00 (0.00) & 0.88 (0.34) & 0.98 (0.15) \\
    9  & 1.00 (0.00) & 1.00 (0.00) & 0.88 (0.33) & 0.92 (0.28) \\
    12 & 0.94 (0.24) & 0.96 (0.21) & 0.81 (0.40) & 0.86 (0.35) \\
    15 & 0.93 (0.27) & 0.98 (0.15) & 0.67 (0.48) & 0.92 (0.28) \\
    \bottomrule
  \end{tabular}
%  \caption{KDE}
  \label{tab: kderegacc}
%  \end{subtable}
  \label{tab: logregacc}
%  \end{subtable}
\end{table}

%$\rem{Again, all time-outs were thrown out for both methods, and only trials where both algorithms finished running were included so as to make a fair comparison without double penalization for MC. 
Table~\ref{tab:accuracy} shows the success rate with respect to the number of data owners under uniform distribution. For every set of parameters except for 3 owners on the Logistic Regression task, SV-Exp outperformed MC. SV-Exp achieves not only higher success rates but also smaller standard deviations. Additionally, it is clear that when the number of data owners increases, MC becomes less and less successful (its success rate with 15 data owners on the Logistic Regression task is only $67\%$). The comparison clearly shows the practical value of SV-Exp.

The difference in performance between the two methods follows from the fact that SV-Exp approximates the \textit{most differential} subset to shift--the returned counterfactual reduces $\Psi(A \setminus \Delta A,B \cup \Delta A')$ as much as possible. The high success rates and low standard deviations in Table~\ref{tab: kderegacc} demonstrate SV-Exp's efficiency and consistency in finding a successful counterfactual.

%The low MC accuracy in Table~\ref{tab: logregacc} speaks to the existence of large counterfactuals between data owners and thus the frequent inability of MC to reach a satisfactory counterfactual in the allotted time. This is also evident in the runtime comparisons in Figure~\ref{fig: runtimecomp}, where the majority of trials that used Logistic Regression as a utility function failed to complete for MC. 

%\subsubsection{Zipfian Distribution}

\begin{figure}[t]
    \centering
    \begin{subfigure}{0.95\linewidth}
        \includegraphics[width=0.95\linewidth]{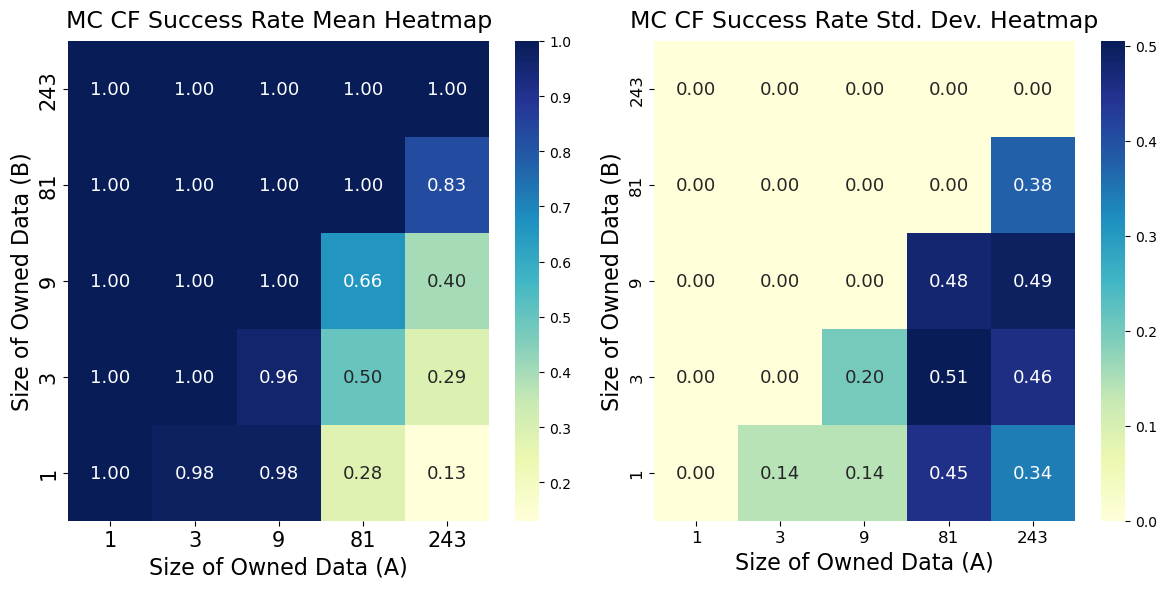}
        \caption{MC Success Rate}
        \label{fig:zipfianlrmcaccuracy}
    \end{subfigure}
    \hfill
    \begin{subfigure}{0.95\linewidth}
        \includegraphics[width=0.95\linewidth]{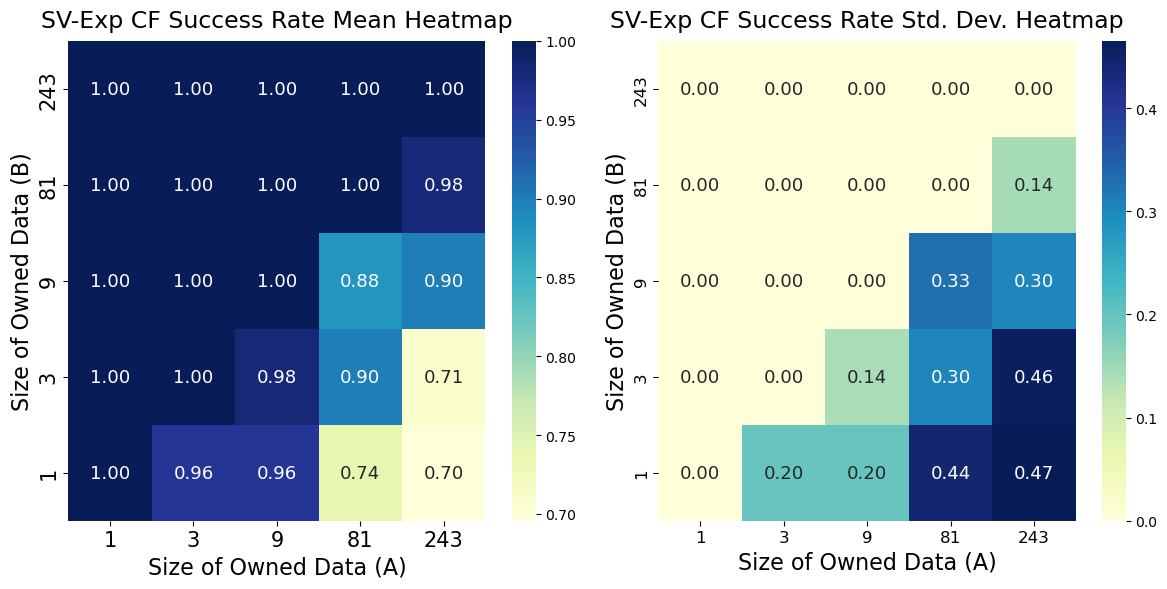}
        \caption{SV-Exp Success Rate}
        \label{fig:zipfianlrgreedyaccuracy}
    \end{subfigure}
    \caption{Statistics of the success rate under the Zipfian distribution of data owners. In each subfigure, the mean is shown in the left and the standard deviation is shown in the right.}
    \label{fig:zipfaccuracy}
\end{figure}

\nop{
\begin{figure*}[t]
    \centering
    \begin{subfigure}{0.48\linewidth}
        \includegraphics[width=\linewidth]{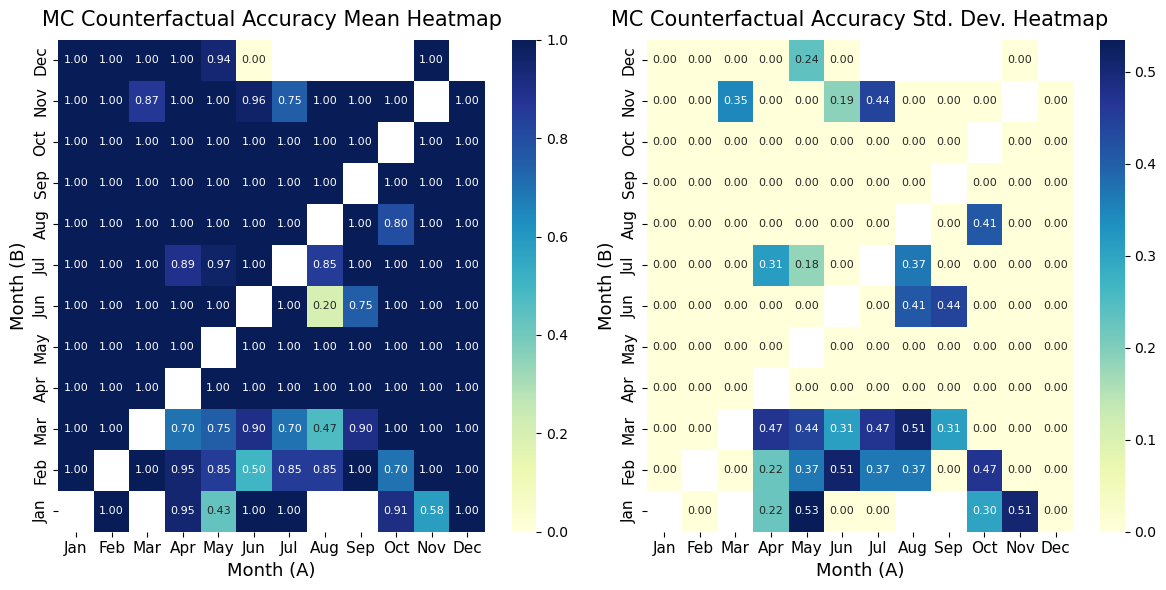}
        \caption{MC Accuracy}
        \label{fig:naturalmcaccuracy}
    \end{subfigure}
    \hfill
    \begin{subfigure}{0.48\linewidth}
        \includegraphics[width=\linewidth]{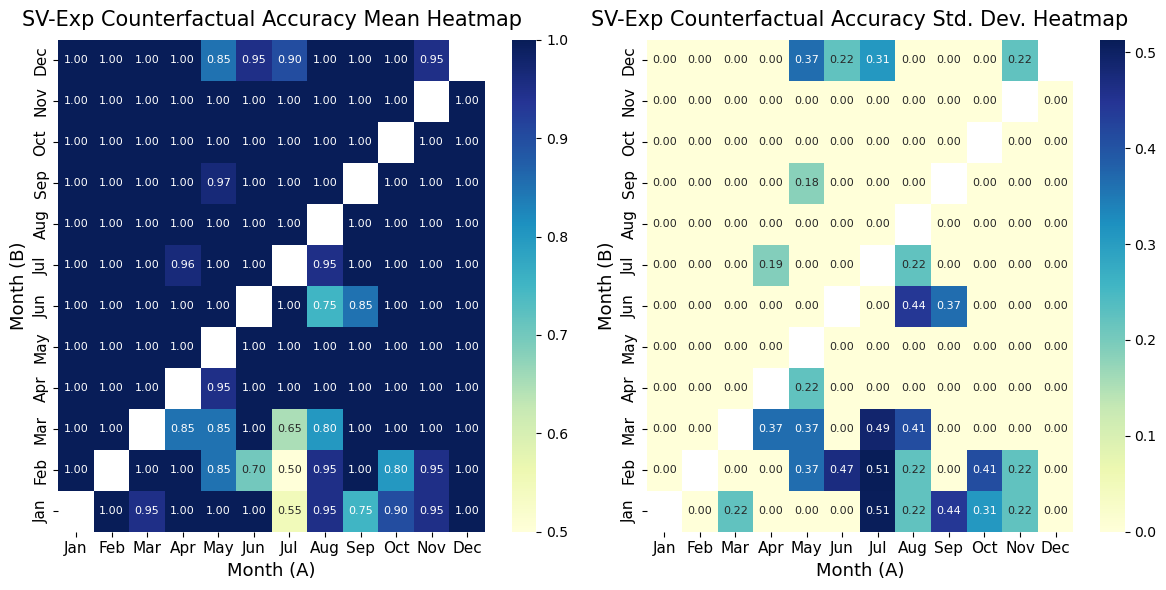}
        \caption{SV-Exp Accuracy}
        \label{fig:naturalgreedyaccuracy}
    \end{subfigure}
    \caption{Statistics of accuracy in the  natural clustering case studies. Blank boxes in the MC heatmap represent cases where every single trial timed out and an average accuracy could not be calculated.}\label{fig:accuracy-naturalclustering}
\end{figure*}
}

Figure~\ref{fig:zipfaccuracy} shows the success rate with respect to data allocation among data owners using the Zipfian distribution. Again, the number of data owners is set to 9.  When the size difference between two data owners becomes large, the problem becomes challenging for both methods as the estimation error may accumulate after many entries are moved to a counterfactual. SV-Exp is still able to achieve a much higher success rate than MC.

\subsection{Case Study 1: Counterfactual Explanations and Feature Selection on the Boston Housing Prices Dataset}\label{sec: boston}

% \begin{figure}[t]
%     \centering
%     \includegraphics[scale=0.3]{linregcoef.png}
%     \caption{Coefficients of a linear regression on the full feature set of the Boston Housing Dataset.}
%     \label{fig:linregcoefall}
% \end{figure}
% \begin{figure}[t]
%     \centering
%     \includegraphics[scale=0.3]{linregselected.png}
%     \caption{Coefficients of a linear regression on four features of the Boston Housing Dataset.}
%     \label{fig:linregselected}
% \end{figure}

In this section, we conduct an interesting case study showing counterfactual explanation can be used as feature selection. In this case study, a dataset is partitioned \emph{vertically} using the Boston Housing Prices dataset--different data owners own different subsets of attributes in a data set. We will conduct another case study where a dataset is partitioned horizontally in Section~\ref{sec: hotel}.
%We motivate this paper with a case study using the Boston Housing Prices dataset, where instead of partitioning the data row-wise, we partition the data feature-wise by assigning each data owner some subset of features. We thus address the reality that the usual separation of data ownership involves features such as location, social statistics, etc. instead of a random uniform distribution of data rows.  

The features of the Boston Housing Dataset include CRIM (per capita crime rate by town), ZN (proportion of residential land zoned for lots over 25,000 sq.ft.), INDUS (proportion of non-retail business acres per town), CHAS (Charles River dummy variable: 1 if tract bounds river, 0 otherwise), NOX (nitric oxides concentration in parts per 10m), RM (average number of rooms per dwelling), AGE (proportion of owner-occupied units built prior to 1940), DIS (weighted distances to five Boston employment centres), RAD (index of accessibility to radial highways), TAX (full-value property-tax rate per \$10k), PTRATIO (pupil-teacher ratio by town), B ($1000(Bk - 0.63)^2$) where Bk is the proportion of blacks by town, LSTAT (\% lower status of the population), and MEDV (Median value of owner-occupied homes in thousands of dollars). MEDV serves as our target variable.

We assign the features to three owners according to their meaning (semantically): (1) property features RM and AGE to owner P; (2) geographic features CHAS, NOX, DIS, and RAD to owner G; and (3) social features CRIM, B, ZN, INDUS, TAX, PTRATIO, LSTAT to owner S. %Data owner 1 will thus own ; owner 2 will own ; and owner 3 will own . This is a reasonable partition we could expect to see in real-life data ownership. We run 50 trials with this structure.

\begin{figure}[t]
    \centering
      \begin{subfigure}
        {0.5\linewidth}
    \includegraphics[scale=0.28]{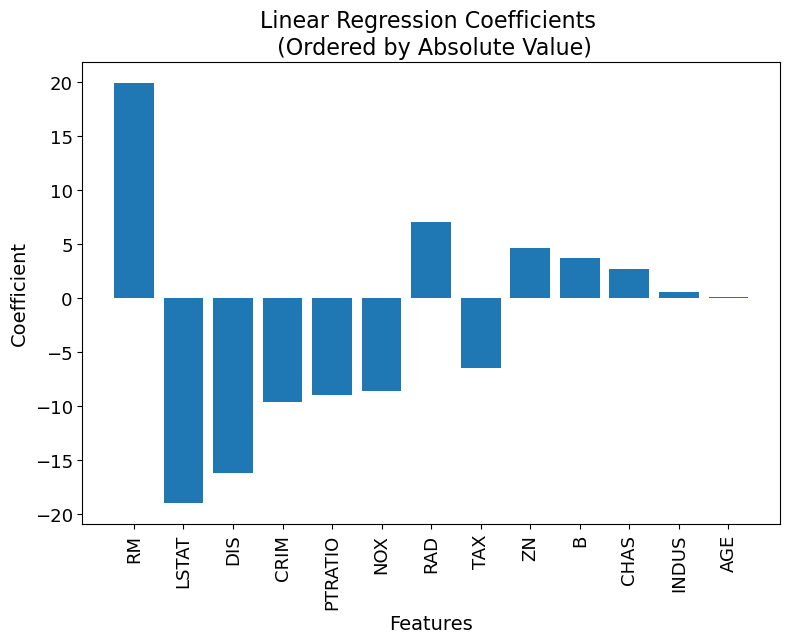}
    \caption{Full feature set}
    \label{fig:linregcoefall}
      \end{subfigure}
      \hfill
      \begin{subfigure}
        {0.3\linewidth}
    \includegraphics[scale=0.28]{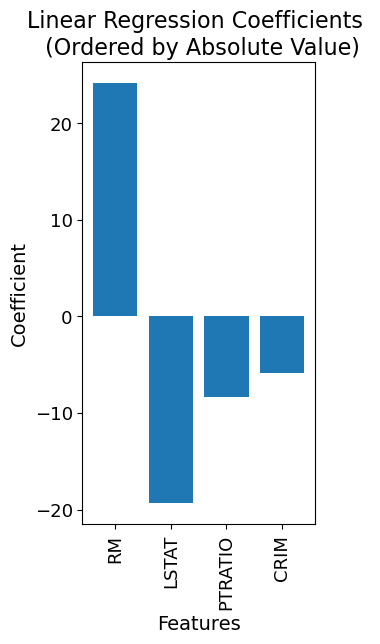}
    \caption{Four features}
    \label{fig:linregselected}
      \end{subfigure}
\caption{Coefficients of a linear regression on the Boston Housing Prices Dataset.}
\label{fig:coeflinreg}
\end{figure}

In this experiment, to keep the discussion easy to understand, we use the entire dataset as our training set and simply use the training error to define our utility function.  Figure~\ref{fig:linregcoefall} shows the coefficients of the attributes in the linear regression on all attributes. We sort all attributes in the coefficient descending order in the figure.

\nop{
This is to prevent uncertainty about the ordering of the Linear Regression coefficients, as using differently sampled training sets may result in alternate orderings and lead to confusion about the interpretation of the counterfactuals.
We now show that importance of the Shapley counterfactual comes from its feature selecting capabilities in cases of collinearity between variables in a Linear Regression task. 
}

Owner S has a higher Shapley value than owner P in the task of linear regression. Which features are in the counterfactual explanation? Since among the social features owned by S, LSTAT and CRIM have the largest absolute values of the coefficients in linear regression, an intuitive guess is that LSTAT and CRIM may be the first two features selected for the counterfactual explanation. Surprisingly, $\{$LSTAT, PTRATIO$\}$ is the counterfactual explanation returned by SV-Exp and does revert the Shapley value relation.

%If we do not use the Shapley counterfactuals and only use information about the coefficients in Figure \ref{fig:linregcoefall}, we would assume that if we take the social features to be owner A and the geographic features to be owner B, if the social features dominate, that the SV-exp greedy counterfactual would be either LSTAT; LSTAT and CRIM; or LSTAT, CRIM, PTRATIO. These are, after all, the features with the largest absolute value coefficients ranked in descending order. But there are cases when the SV-Exp algorithm denotes LSTAT and PTRATIO to be the counterfactual, which seems to be a contradiction of the coefficient ordering of the social features.

We further investigate the correlation among several features as shown in Table~\ref{tab: corrmat}. Interestingly, CRIM is more correlated with LSTAT than PTRATIO is. This means, after LSTAT is chosen in the counterfactual explanation, choosing CRIM is not as effective as choosing PTRATIO in further reducing the utility of owner S and enhancing the utility of owner P. Indeed, if we conduct a linear regression using only four features, RM, which is the dominating feature owned by P, LSTAT, PTRATIO, and CRIM, the absolute value of coefficient of PTRATIO is larger than that of CRIM, as shown in Figure~\ref{fig:linregselected}.

% Exploring further, we see that the SV-Exp algorithm has in effect done an automatic feature selection that accounts for multicollinearity—in the presence (in the same data partition) of one of the strongest features LSTAT, CRIM is more correlated with LSTAT than PTRATIO, as shown in Table \ref{tab: corrmat}. This makes PTRATIO the more logical counterfactual choice when moving LSTAT itself does not successfully flip the Shapley. This choice is confirmed and consistent with the coefficient rankings shown in Figure \ref{fig:linregselected}. 
Essentially, SV-Exp automatically feature selects one of the variables when there are different levels of correlation, simplifying how we deal with multicollinearity and understand the features of datasets with one algorithm. We make a case for the Shapley counterfactual as a tool for feature selection.

In this illustrative case study, the counterfactual explanation of Shapley values not only shows the comparative advantages between two feature sets but also facilitates the identification and enhancement of features within one set based on those of the other.

\begin{table}[t]
\centering
\caption{Correlation matrix between four selected features in the Boston Housing Prices dataset.}
\begin{tabular}{rrrrr}
\toprule & \textbf{LSTAT} & \textbf{CRIM} & \textbf{RM} & \textbf{PTRATIO} \\
\midrule
\textbf{LSTAT} & 1.00 & 0.46 & -0.61 & 0.37 \\
\textbf{CRIM} & 0.46 & 1.00 & -0.22 & 0.29 \\
\textbf{RM} & -0.61 & -0.22 & 1.00 & -0.36 \\
\textbf{PTRATIO} & 0.37 & 0.29 & -0.36 & 1.00 \\
\bottomrule
\end{tabular}
 %\vspace{-5mm}
\label{tab: corrmat}
\end{table}

\subsection{Case Study 2: Counterfactual Explanations and Differences of Distributions on the Hotel Reservations Dataset}\label{sec: hotel}

In this case study, we partition the Hotel Reservation dataset \emph{horizontally}.
%We construct a case study using the Hotel Reservations dataset to further motivate the value of Shapley value explanability. 
In this experiment, we use logistic regression to predict the probability of whether a client will cancel the reservation. We assign to each data owner all data of one month and thus we have in total 12 data owners. The research question aims to determine which months' data contribute more in modeling reservation cancellations. By computing the counterfactual explanations we are interested in understanding the differences in contribution between two months.
%This type of categorical assignment will introduce another way we can use counterfactual explanations to understand the Shapley value in the context of categorical variables and the contribution of each category.  
We perform pairwise experiments, testing all 144 possible pairs of months and thus data owners over 10 trials per pairing. The diagonals represented the cases where $A$ and $B$ are the same data owner and were thus moot experimentally. The sizes of the data subsets from January to December owned by the 12 data owners are 25, 37, 52, 45, 59, 61, 65, 79, 105, 117, 73, and 82, respectively. The Shapley value of each month's data is detailed in Table~\ref{tab: month shapleys}.

\begin{table}[t]
    \centering
    \caption{Shapley values of the month's data estimated using Monte Carlo Sampling, where utility $=\eta-$log-loss, $\eta=20$, and the task is Logistic Regression.}
    \begin{tabular}{cccccc}
    \textbf{Jan}&\textbf{Feb} & \textbf{Mar} & \textbf{Apr} & \textbf{May} & \textbf{Jun}  \\
         0.71&1.58 & 1.67 & 1.78 & 1.72 & 1.87  \\
         \midrule
    \textbf{Jul} & \textbf{Aug} & \textbf{Sep} & \textbf{Oct} & \textbf{Nov} & \textbf{Dec} \\
    1.86 & 1.84 & 1.79 & 1.80 & 1.83 & 1.79\\
    \end{tabular}
    % \vspace{-8mm}
    \label{tab: month shapleys}
\end{table}

\begin{figure}[t]
    \centering
    \includegraphics[scale=0.43]{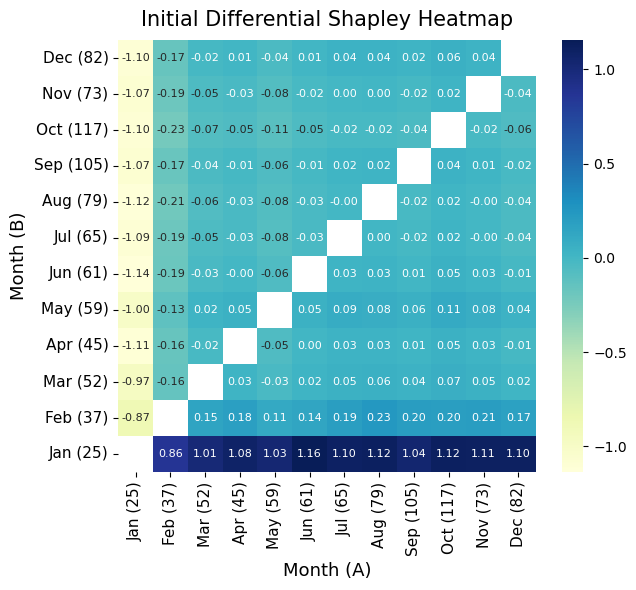}
    \caption{Initial Differential Shapley ($\Psi_{\mathbb{O}}(A,B)$) Between Data Owners.}
    \label{fig:initdiffhotel}
\end{figure}

Figure \ref{fig:initdiffhotel} shows the initial difference in Shapley value between different months. Theoretically the matrix should be symmetric. However, due to the estimation errors in practice, the matrix is not perfectly so.

\begin{figure}[t]
\includegraphics[scale=0.3]{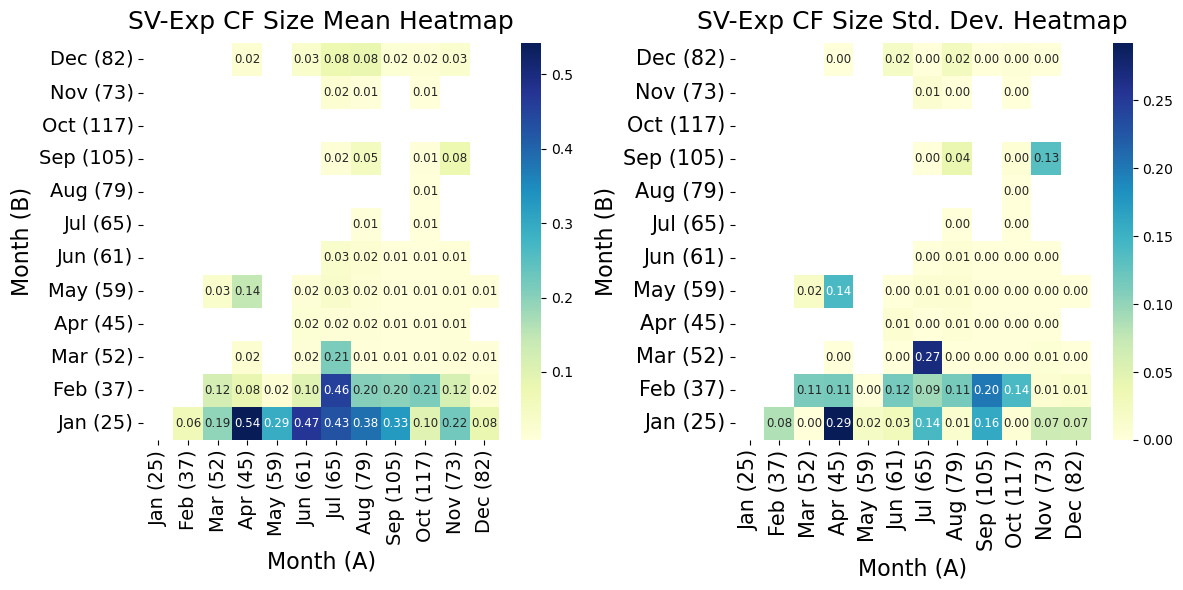}
\caption{Average size of counterfactuals (from the SV-Exp algorithm) between data owners in percentage of $A$'s original data. Zero values are left blank for more visual clarity, and sizes of each month's data are in parentheses on the axes labels.}
\label{fig:hotelcfsize}
\end{figure}

The differential Shapley values between all months and January are similar, as shown in Figure~\ref{fig:initdiffhotel}, but the sizes of the Shapley counterfactuals are vastly different (Figure \ref{fig:hotelcfsize}). This is a clear example showing that the similar differential Shapley values may mask important differences about how advantages between pairs of data owners may be established. Take March and April as a concrete example. Both of their Shapley values are about $1.0$ larger than the Shapley value of January. Moreover, their dataset sizes are also close, March having 52 data entries and April 45. But the counterfactual explanation of March against January takes about $19\%$ of the data in March, while the counterfactual explanation of April against January takes $54\%$ of the data in April. This indicates that the data in April is much more ``robust'' than that in March with respect to the data in January—most likely, the data distribution of April is more different than that of January, while the data distribution in March is more similar. To verify if this insight holds, we use the Wasserstein distance between two months to understand their relative differences in distribution. The Wassertein distance, or the earth mover's distance, measures the estimated ``cost'' of turning one probability distribution into the other. The Wasserstein distance between March and January is 3.24, while the Wasserstein distance between April and January is 6.52.  The much larger Wasserstein distance between April and January confirms our conjecture. Just for the reader's reference, the Wasserstein distance between March and April is 5.59.

This case study shows another interesting use of the counterfactual explanation of Shapley values, capturing and understanding the differences of data distributions between pairs of data owners.

\begin{figure}[t]
    \centering
    \includegraphics[scale=0.43]{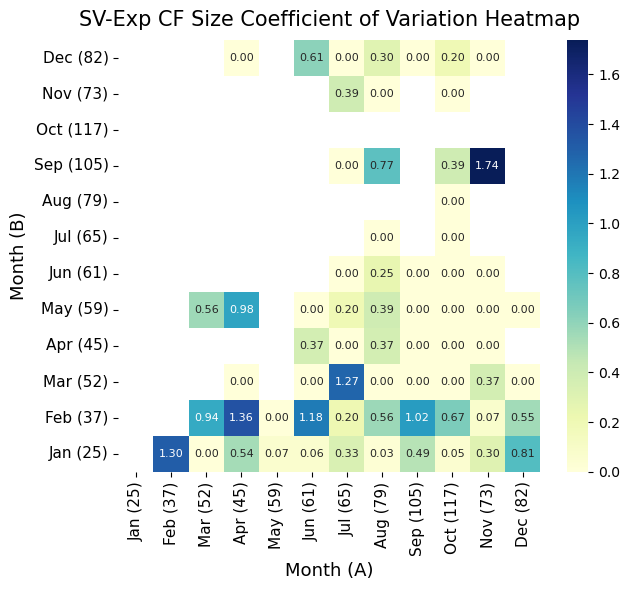}
    \caption{Coefficient of Variation on the average size of counterfactuals produced by SV-Exp. Coefficient of Variation is the ratio of the standard deviation to the mean.}
    \label{fig:coefofvar}
\end{figure}

%Additionally, see that July (as data owner $A$) has about the same initial differential Shapley value's against January, February, and March as all the other months (including September and October, who both own much more data than July), but it tends to have much higher counterfactuals against January, February, and March, speaking to its own robustness as a dataset—another fact hidden by knowing only the differential Shapleys.

%\todo{interpretation of coefficient of varation graph}

\subsection{Stability of Approximate Counterfactual Explanations in SV-Exp}
 
%We now focus on the larger counterfactuals, specifically in the bottom two rows, for further analysis (see Figure \ref{fig:initdiffhotel}). This is due to the fact that in cases where $|\Psi_{\mathbb{O}}(A,B)|$ is small, the Shapley can be flipped by one or two items, which does not provide much insight into the meaning of the Shapley value. Thus, our method is best suited to when multiple owners have significantly different Shapleys—by comparing the counterfactuals amongst them, we can glean more information about the robustness of each owner's data. 

SV-Exp approximates counterfactual explanations of Shapley values. We now explore one last question: given that SV-Exp draws random samples of coalitions, how stable are the approximate counterfactuals? In this question, we only consider those cases where the differential Shapley values are large, since a small differential Shapley value can be reverted easily using any small counterfactual. For example, if the initial differential Shapley is only very slightly positive and any random item from $A$ could flip the Shapley relation, we do not get much more information about $A$ and $B$ from the counterfactual. Thus, since we will be studying the content of the outputted SV-Exp counterfactuals, we focus on those that are larger—in these cases, the data entries making up the counterfactuals will have some level of significance. We still use the Hotel Reservations dataset as in Section~\ref{sec: hotel}.

We look at the \emph{coefficient of variation}, also known as normalized root-mean-square deviation (NRMSD), percent RMS, and relative standard deviation (RSD), which is defined as the ratio of the standard deviation over the mean. Interestingly, in Figure \ref{fig:coefofvar}, there are many cases where the coefficient of variation is small, such as March, May, June, August, and October against January. In those cases, SV-Exp often chooses the same subset of data entries in many trials. As a concrete example, in the case of the counterfactual of May against January, the subset of 18 data entries in May, $\{$162, 628, 711, 651, 69, 63, 678, 379, 281, 524, 74, 355, 406, 439, 758, 662, 615, 609$\}$ (the numbers of the record-ids in the dataset) is chosen every time by SV-Exp. This shows that SV-Exp demonstrates surprising consistency on some datasets.

% This means that despite the amount of different coalitions that can occur ($2^{11}$ possible subsets of the 11 other data owners), through sampling, those 18 items consistently perform the best. 

There are also some cases where the coefficient of variation is large. For example, April has a similar Shapley value as March (Table~\ref{tab: month shapleys} and Figure~\ref{fig:initdiffhotel}), but its coefficient of variation is high as it has a large standard deviation. Looking at the detailed counterfactual results, we find that counterfactual explanations of April against January can be as small as size 13 ($\{$528, 289, 312, 218, 204, 725, 398, 559, 671, 301, 710, 774, 572$\}$) and as large as size 40 ($\{$572, 33, 462, 312, 588, 692, 792, 475, 559, 344, 192, 785, 301, 155, 134, 473, 360, 725, 671, 501, 218, 502, 362, 184, 248, 142, 141, 292, 461, 185, 510, 289, 81, 528, 95, 774, 346, 204, 398, 710$\}$). Moreover, SV-Exp implicitly ranks the data entries from most to least power in the subset, the ranking of the entries is also very different from run to run. This indicates that entries in the data of April may have utilities that are much more sensitive to what coalition the samples drawn.
%Thus, we show that when the initial differential Shapley is large, comparing the counterfactuals between different pairs of data owners can tell us a lot about their original Shapley—ie, how their data entries contribute to overall performance in the presence of coalitions.}
%\vspace{4mm}

\section{Related Work}\label{sec:related}

In Section~\ref{sec:intro}, we already discuss a series of related work on data markets, the Shapley value, and computation. To the best of our knowledge, we are the first to formulate the problem of counterfactual explanation of the Shapley value.  In this section, we focus on briefly reviewing the related work on counterfactual explanations.

Counterfactual explanations~\cite{wachter2017counterfactual, sokol2019counterfactual, moraffah2020causal, akula2020cocox, fong2017interpretable, grace} have been widely used in interpreting and understanding algorithmic decisions made in many real world applications~\cite{akula2020cocox, fong2017interpretable, grace}. 
Those methods~\cite{fong2017interpretable, akula2020cocox, van2019interpretable, moraffah2020causal} often explain a prediction on a given case using small and interpretable perturbations on the case such that the prediction is changed~\cite{moraffah2020causal}. For example, \citet{fong2017interpretable} interpret image prediction by identifying the smallest pixel-deletion mask that causes the most significant drop of the prediction score. 
\citet{akula2020cocox} find image patches that need to be added to or deleted from an input image in order to change the prediction. \citet{van2019interpretable} use class prototypes to produce counterfactual explanations that are close to the classifier's training data distribution.
\citet{moore2019explaining} propose a method to generate counterfactual explanations from adversarial examples with gradient constraints.
\citet{grace} propose an entropy-based feature selection method to limit the features to be perturbed. \citet{10.14778/3461535.3461546} compute understandable counterfactual explanations for Kolmogorov-Smirnov Test results. \citet{NEURIPS2021_2c8c3a57} provide counterfactual explanations for graph nueral networks.

Surprisingly, the problem of a counterfactual explanation of the Shapley value has not been studied in literature. Moreover, the existing counterfactual explanation methods cannot be applied to explain the Shapley value directly.

%, as we do in our model where two data owners are involved in the comparison.
%\vspace{-4mm}
\section{Limitations and Extensions}\label{sec:ext}

Counterfactual explanations of the Shapley value present a novel and significant challenge. This study represents an initial foray into this promising problem. Despite our encouraging advancements, our SV-Exp method remains subject to certain limitations. Chief among these constraints is its scalability. Although SV-Exp demonstrates superior scalability compared to brute force and Monte Carlo methods, it encounters difficulties when confronted with very large datasets. This challenge arises primarily due to the extensive training requirements inherent in SV-Exp. Notably, the utility function, utilized twice for every sample, necessitates the training of a new model for each iteration. Consequently, when handling substantial datasets or data necessitating convolutional neural networks (CNNs) or more intricate training algorithms, SV-Exp may prove inefficient in identifying counterfactual explanations.

Our SV-Exp method can also account for variances of the counterfactual explanation problem for Shapley values. For example, instead of transferring items from $A$ to $B$, what is the minimal number of items we need to simply delete from $A$ in order for $\phi(A)<\phi(B)$? Essentially, what is the minimal set of entries in $A$ that make up the difference in Shapley value between $A$ and $B$? This is still an NP-hard problem with a feasible solution (worst case, delete all of $A$), and our algorithm can easily be extended to approximate a solution for this problem by simply taking away the step where we transfer data entries to $B$. %This is an interesting and useful question because approximating a solution will give us another type of explanation for how much $A$ truly contributes more than $B$ in a data marketplace, something that knowing only the absolute Shapley values does not illuminate.

%\todo{discuss future work like Shapley p-val?}

%\vspace{4mm}
\section{Conclusion}\label{sec:con}

Data valuation is a fundamental mechanism within a data market. While there are more and more studies on efficient data valuation, how we understand and explain data valuation remains an open problem. In this paper, we formulate the problem of counterfactual explanation of the Shapley value in data coalitions, which, to the best of our knowledge, is the first study tackling this important issue. We show the complexity of the problem, propose a series of techniques, and develop a greedy approximation method.  Our experiments on real datasets clearly show the efficiency of our approach and the effectiveness of a counterfactual Shapley explanation in interpreting data value, feature selection, and detecting data distribution differences.

Our study
%give one clear metric for evaluating the ``quality'' of data according to the Shapley value and 
illuminates a methodical way to interpret how individual data entries contribute to the overall value of a data owner's dataset in a game theoretic setting. This paper opens a new direction for promising future work. For example, building on the general framework developed in this study, it is interesting to explore more effective and efficient approaches to Shapley value interpretation in specific types of data collaboration, such as data assemblage~\cite{10.14778/3551793.3551829, SIGMOD24:shapley-value-simple-game}. Moreover, in some applications, one may be interested in finding coherent subsets of data as counterfactual explanations. It will also be important to explore the explanation of other types of data valuation metrics for data markets.

\begin{acks}
We thank Kiran Dwivedi and Phillip Si for their invaluable comments and feedback, as well as Eric Zelikman and Kate Donahue for useful discussions. This research is supported in part by a startup grant and a Beyond the Horizon grant by Duke University. All opinions, findings, conclusions and recommendations in this paper are those of the authors and do not necessarily reflect the views of the funding agency.
\end{acks}

%\begin{acks}
% This work was supported by the [...] Research Fund of [...] (Number [...]). Additional funding was provided by [...] and [...]. We also thank [...] for contributing [...].
%\end{acks}

%\clearpage

\bibliographystyle{ACM-Reference-Format}
\bibliography{ref,datapricingref,counterfactual}

\end{sloppy}
\end{document}